\documentclass[11pt]{article}

\usepackage{latexsym,amsmath,amssymb,amscd,eucal}
\usepackage{xypic}
\usepackage{enumerate}

\def\harr#1#2{\smash{\mathop{\hbox to .3in{\rightarrowfill}}
 \limits^{\scriptstyle#1}_{\scriptstyle#2}}}

\def\appendix#1{\addtocounter{section}{1}\setcounter{equation}{0}
\renewcommand{\thesection}{\Alph{section}}
\section*{Appendix \thesection\protect\indent \parbox[t]{11.715cm} {#1}}
\addcontentsline{toc}{section}{Appendix \thesection\ \ \ #1} }
\newcommand{\eq}{\begin{equation}}
\newcommand{\eqend}{\end{equation}}

\newbox\ncintdbox \newbox\ncinttbox
\setbox0=\hbox{$-$} \setbox2=\hbox{$\displaystyle\int$}
\setbox\ncintdbox=\hbox{\rlap{\hbox to
\wd2{\hskip-.125em\box2\relax \hfil}}\box0\kern.1em}
\setbox0=\hbox{$\vcenter{\hrule width 4pt}$}
\setbox2=\hbox{$\textstyle\int$}
\setbox\ncinttbox=\hbox{\rlap{\hbox to
\wd2{\hskip-.175em\box2\relax \hfil}}\box0\kern.1em}

\newcommand{\torus}{{\mathbb T}}
\def\be{\begin{equation}}
\def\ee{\end{equation}}
\def\bea{\begin{eqnarray}}
\def\eea{\end{eqnarray}}
\def\bd{\begin{displaymath}}
\def\ed{\end{displaymath}}

\DeclareFontFamily{U}{rsf}{}
\DeclareFontShape{U}{rsf}{m}{n}{
  <5> <6> rsfs5 <7> <8> <9> rsfs7 <10-> rsfs10}{}
\DeclareMathAlphabet\Scr{U}{rsf}{m}{n}

\def\cR{{\Scr R}}

\def\cC{{\Scr C}}
\def\cK{{\Scr K}}
\def\cD{{\Scr D}}

\def\cP{{\Scr P}}
\def\cT{{\Scr T}}

\def\cE{{\Scr E}}
\def\cF{{\Scr F}}

\def\cG{{\Scr G}}
\def\cO{{\Scr O}}

\makeatletter
\newdimen\normalarrayskip              % skip between lines
\newdimen\minarrayskip                 % minimal skip between lines
\normalarrayskip\baselineskip
\minarrayskip\jot
\newif\ifold             \oldtrue            
\def\arraymode{\ifold\relax\else\displaystyle\fi} % mode of array entries
\def\@arrayskip{\ifold\baselineskip\z@\lineskip\z@
     \else
     \baselineskip\minarrayskip\lineskip2\minarrayskip\fi}
\def\@arrayclassz{\ifcase \@lastchclass \@acolampacol \or
\@ampacol \or \or \or \@addamp \or
   \@acolampacol \or \@firstampfalse \@acol \fi
\edef\@preamble{\@preamble
  \ifcase \@chnum
     \hfil$\relax\arraymode\@sharp$\hfil
     \or $\relax\arraymode\@sharp$\hfil
     \or \hfil$\relax\arraymode\@sharp$\fi}}
\def\@array[#1]#2{\setbox\@arstrutbox=\hbox{\vrule
     height\arraystretch \ht\strutbox
     depth\arraystretch \dp\strutbox
     width\z@}\@mkpream{#2}\edef\@preamble{\halign \noexpand\@halignto
\bgroup \tabskip\z@ \@arstrut \@preamble \tabskip\z@ \cr}%
\let\@startpbox\@@startpbox \let\@endpbox\@@endpbox
  \if #1t\vtop \else \if#1b\vbox \else \vcenter \fi\fi
  \bgroup \let\par\relax
  \let\@sharp##\let\protect\relax
  \@arrayskip\@preamble}
\makeatother

\newcommand{\beq}{\begin{eqnarray}}
\newcommand{\eeq}{\end{eqnarray}}

\newcommand{\G}{\Gamma}

\def\appendix#1{\addtocounter{section}{1}\setcounter{equation}{0}
\renewcommand{\thesection}{\Alph{section}}
\section*{Appendix \thesection. #1}
\addcontentsline{toc}{section}{Appendix \thesection\ \ \ #1} }

\newtheorem{theorem}{Theorem}[section]

\newtheorem{corollary}{Corollary}[section]
\newtheorem{proposition}{Proposition}[section]

\newtheorem{conjecture}{Conjecture}[section]

\newtheorem{definition}{Definition}[section]

\newtheorem{remark}{Remark}[section]

\numberwithin{equation}{section}

\begin{document}

\begin{flushright}
YITP - 07 - 53
%SISSA/nn/06/EP\\ hep-th/0602162n
\end{flushright}

\vspace{.1in}

\begin{center}
{\Large\bf HOMOLOGY AND K--THEORY METHODS FOR CLASSES OF BRANES WRAPPING NONTRIVIAL CYCLES}

\end{center}
\vspace{0.1in}
\begin{center}
{\large 
%L. Bonora $^{(a)}$\footnote{ bonora@sissa.it}
%and 
A. A. Bytsenko $^{(a)}$ $^{(b)}$
\footnote{abyts@uel.br}} 
%and M. E. X.
%Guimar\~aes $^{(c)}$ \footnote{marg@unb.br}} 
\vspace{7mm}
\\
%$^{(a)}$ {\it International School for Advanced Studies (SISSA/ISAS) \\
%Via Beirut 2, 34014 Trieste, Italy  and INFN, Sezione di
%Trieste} \vspace{5mm}\\
$^{(a)}$ {\it
Departamento de F\'{\i}sica, Universidade Estadual de
Londrina\\
Caixa Postal 6001, Londrina-Paran\'a, Brazil}\\
$^{(b)}$ {\it Yukawa Institute for Theoretical Physics, Kyoto University, Kyoto 606-8502, Japan}
%\vspace{5mm}\\
%$^{(c)}$ {\it Departamento de Matem\'atica, %Universidade
%de Bras\'{\i}lia, DF, Brazil}\\
\end{center}
\vspace{0.1in}
%~\\
\begin{center}
{\bf Abstract}
\end{center}
We apply some methods of homology and K-theory to special classes of branes wrapping homologically nontrivial cycles. We treat 
the classification of four-geometries in terms of compact  stabilizers (by analogy with Thurston's classification of three-geometries) and derive the K-amenability of Lie groups associated with locally symmetric spaces listed in this case. 
More complicated examples of T-duality and topology change from fluxes are also considered. We analyse D-branes and fluxes in type II string theory on ${\mathbb C}P^3\times \Sigma_g \times {\mathbb T}^2$ with torsion $H-$flux and demonstrate in details  the conjectured T-duality to ${\mathbb R}P^7\times X^3$ with no flux. In the simple case of $X^3 = {\mathbb T}^3$, T-dualizing the circles reduces to duality between
${\mathbb C}P^3\times {\mathbb T}^2 \times {\mathbb T}^2$ with $H-$flux and  ${\mathbb R}P^7\times 
{\mathbb T}^3$ with no flux.

\vfill


%\maketitle

\newpage
\tableofcontents
%\vfil\eject

\newpage

\section{Introduction}

The problem of classifying geometries is one of the main problem in complex analysis and in mathematics as a whole, and plays a fundamental role in physical models.
Every one-dimensional manifold is either $S^1$ (closed, 
i.e. compact with empty boundary) or ${\mathbb R}$ (open), with a unique topological, piecewise linear and smooth structure and orientation. 
All curves of genus zero can be uniformized by rational functions, all those of genus one can be uniformized by elliptic functions,
and all those of genus more than one can be uniformized by
meromorphic functions, defined on proper open subsets of ${\mathbb C}$. This result, due to Klein, Poincar\'e and Koebe, is one of the deepest achievements in mathematics. 
A complete solution of the uniformization problem has not yet been obtained, excepted for the one-dimensional complex case. However, there were advances in this problem, which have been
essential to understand the foundations of topological methods, covering spaces, existence theorems for partial differential equations, existence and distorsion theorems for conformal mappings, etc.

In accordance with Klein-Poincar\'e uniformization theorem, each Riemann surface can be represented (within a conformal
equivalence) in the form $\Sigma/\Gamma$, where $\Sigma$ is one of the three canonical regions, namely: The extended plane
$\bar{\mathbb C}$ (the sphere ${\mathbb S}^2$), the plane
${\mathbb C}$ (${\mathbb R}^2$), or the disk, 
and $\Gamma$ is a
discrete group of M\"{o}bius automorphisms of $\Sigma$ acting
freely there. Riemann surfaces with such coverings are elliptic,
parabolic and hyperbolic type, respectively. This theorem admits a
generalization also to surfaces with branching. 
%A different
%approach to the solution of the uniformization %problem was
%proposed by Koebe. The general uniformization %principle of Koebe asserts that if a Riemann %surface $\widetilde{\Sigma}$ is topologically %equivalent to a planar region 
%${P}$, then
%there also exists a conformal homeomorphism of
%$\widetilde{\Sigma}$ onto ${P}$. The same %problem of
%analytic uniformization reduces to the %topological problem of
%finding all the (branched, in general) planar %coverings
%$\widetilde{\Sigma} \rightarrow \Sigma$ of a %given Riemann surface
%$\Sigma$. The solution of this topological %problem is given by the
%theorem of Maskit. 
With the help of standard uniformization
theorems and decomposition theorems \cite{Maskit}, one can
construct and describe all the uniformizations of Riemann surfaces
by Kleinian groups. Furthermore, by using the quasiconformal
mappings, one can obtain an uniformization theorem of more general character
\footnote{This fact is related to Techm\"{u}ller
spaces.}. 
Namely, it is possible to prove that several surfaces can be uniformized simultaneously. 
For any closed orientable two-dimensional manifold
$\Sigma_{\Gamma}$ the following result holds:
Every conformal structure on
$\Sigma_{\Gamma}$ is represented by a constant curvature
geometry. The only
simply connected manifolds with constant curvature are
${\mathbb S}^2$ or
${\mathbb R}^2$ or ${\mathbb H}^2$ and
$\Sigma_{\Gamma}$ can be represented as
$\Sigma/\Gamma$, where $\Gamma$ is a group of isometries.

An important progress in the three-dimensional case has been made by
Thurston \cite{Thurston}. 
To analyse this case we should consider the eight geometric structure in the classification of three-manifolds introduced by Thurston.
We also note the recent asserts for Ricci flow on class of three-manifolds. The Ricci flow with surgery was considered in \cite{Hamilton}. A canonical Ricci flow defined on largest possible subset of spacetime has been constructed in \cite{Perelman1,Perelman2}.

In this paper we will analyse in detail the K-theory groups which are the more appropriated arena to classify the D-branes wrapping topologically nontrivial manifolds.
We contemplate various three- and four-geometries giving the relevant meaning of low dimensional brane cycles and describing some new results on K-amenability in the list of four-geometries. We concentrate our analysis on application of homological and K-theory methods to branes and Ramond-Ramond (RR) flux in type II string theory. The RR fields are typically supported on D-branes and take values in appropriate K-theory groups
\cite{Minasian}. These facts have prompted intensive investigations in both the mathematical and physical literature into the properties and definitions of various K-theory groups. 

This paper is outlined as follows. The case of two-geometries is shortly discussed in Section 2.1 and then in Section 2.2 we present the Thurston's list of three-geometries. This list has been organized in terms of the compact stabilizers $\Gamma_x$ of $x\in X$ isomorphic to ${\bf SO}(3)$, ${\bf SO}(2)$ or trivial group $\{1\}$. The analogue list for
four-geometries and the corresponding stabilizer-subgroups of the kind 
${\bf SO}(4)$,\,
${\bf U}(2)$,\,
${\bf SO}(2)\times {\bf SO}(2)$,\,
${\bf SO}(3)$,\,
${\bf SO}(2)$,\,
${\bf S}^1$,\, and $\{1\}$
are analysed in Section 2.3.

The required mathematical tools considered in the paper are exposed in the Section 3. We also refer the reader to the  Appendix where the necessary material on Eilenberg-MacLane spaces is presented.
The Kasparov's KK-pairings and the concept of K-amenable
groups are considered in the Section 3.1 and the Section 3.2, respectively. 

The different aspects of K-group theory are formulated in the
Section 4.
Twisted crossed product of $C^*-$algebras and  twisted K-groups are considered in the Section 4.1.

Remind that some general theorems about the K-groups of $C^*-$algebras have been obtained in \cite{Elliott,Bellissard,Connes},
while perfect analysis of Baum-Connes type conjectures on the K-theories of twisted $C^*-$algebras was considered in \cite{Marcolli,Mathai}. In the special case of general statements,
$
K_*(C^*(\mathbb Z^n,\sigma)) \cong K_*(C^*(\mathbb Z^n)) \cong K^*(\mathbb T^n)
$
for any multiplier (i.e. group two-cocycle) $\sigma$ on $\mathbb Z^n$. The twisted group $C^*-$algebras $C^*(\mathbb Z^n,\sigma)$ have been called {\it noncommutative tori}.
In \cite{Packer,Packer2} this calculation was generalized for
K-groups of the twisted group $C^*-$algebras of uniform lattices in solvable groups. Namely, if $\Gamma$ is a uniform lattice in a solvable Lie group $G$, then
\begin{equation}
K_*(C^*(\Gamma,\sigma)) \cong K^{* + \dim G}(\Gamma\backslash G,
\delta(B_\sigma))\,,
\nonumber
\end{equation}
where $\sigma$ is any multiplier
on $\Gamma$, $K^{*}(\Gamma\backslash G,
\delta(B_\sigma)) $ denotes the twisted K-theory (see for detail \cite{Rosenberg}) of a continuous trace $C^*-$algebra $B_\sigma$ with spectrum $\Gamma\backslash G$, and $\delta(B_\sigma)
\in H^3(\Gamma\backslash G, \mathbb Z)$ denotes the
Dixmier-Douady invariant of $B_\sigma$. This result has been prooved by using the Packer-Raeburn stabilization trick \cite{Packer} and the Thom isomorphism theorem for the K-theory of $C^*$ algebras. Then in \cite{Carey} the main theorem of \cite{Packer,Packer2} has been extended to the case when
$\Gamma$ is a lattice in a K-amenable Lie group $G$ \cite{Kasparov}. For such $G$ and $\Gamma$,
\begin{equation}
K_*(C^*(\Gamma,\sigma))\cong K_*(C^*_r(\Gamma,\sigma))\,,\,\,\,\,\,\,\,\,
K_*(C^*(\Gamma,\sigma)) \cong K^{* + {\rm dim}(G/K)}(\Gamma\backslash G/K, \delta(B_\sigma))\,,
\nonumber
\end{equation}
where $K$ is a maximal compact subgroup of $G$, $\sigma$ is any multiplier on $\Gamma$, $K^{*}(\Gamma\backslash G/K,
\delta(B_\sigma))$ is the twisted K-theory of a continuous trace $C^*-$algebra $B_\sigma$ with spectrum $\Gamma\backslash G/K$.

For the three-dimensional case these results are described in the Section 4.2. 
Then, in the Section 4.3 we proof the K-amenability for a set of Lie groups associated with locally symmetric spaces listed in the four-dimensional case. We formulate the following statement
in the form of conjecture:  If $K(\Gamma, 1)$ is a connected, compact, four-dimensional manifold which is an Eilenberg-MacLane space with fundamental group $\Gamma$, then for any multiplier $\sigma\in H^2(\Gamma, U(1))$ on $\Gamma$ 
\begin{equation}
K_j(C_r^*(\Gamma,\sigma)) \cong K_j(C_r^*(\Gamma)) \cong K^{j+1}(K(\Gamma, 1)), \quad j=0,1.
\nonumber
\end{equation}

The properties of the T-duality for principal higher rank torus bundles with H-flux have been studied in \cite{Bouwknegt}.
This construction of duality can be argued from the so-called generalized Gysin sequence. T-duality for torus bundles comes with the isomorphism of twisted cohomology and twisted cyclic homology, it was analysed via noncommutative topology in 
\cite{Mathai1,Mathai2}. It has been argued \cite{Bouwknegt04,Mathai3} that for type II string theory a general principal torus bundle with general H-flux is a bundle of noncommutative, nonassociative tori. Useful theorems to perform the T-duality computations for circle bundles with H-fluxes are given in \cite{Bouwknegt1,Bouwknegt4}.
In the Section 5 we discuss some more complicated examples of T-duality for circle bundles and topology change from fluxes, taking into account methods of homology and K-theory. We apply to D-branes classification and fluxes in type II string theory on ${\mathbb C}P^3\times \Sigma_g \times {\mathbb T}^2$ with torsion $H-$flux (similar model has been considered in \cite{Bouwknegt2,Mathai3})
and demonstrate in details  the conjectured T-duality to ${\mathbb R}P^7\times X^3$ with no flux. In the simple case of $X^3 = {\mathbb T}^3$, T-dualizing the circles reduces to duality between
${\mathbb C}P^3\times {\mathbb T}^2 \times {\mathbb T}^2$ with $H-$flux and  ${\mathbb R}P^7\times 
{\mathbb T}^3$ with no flux.

\section{Low-dimensional cycles}

\subsection{Two-geometries}
Two-manifolds are complete classified: their piecewise linear and smooth structures are unique and depend only on the homeomorphic type. The homeomorphism type is determined by their fundamental group. If $X$ is a closed orientable two-manifold, then it is either ${\mathbb S}^2, {\mathbb T}^2 = S^1\times S^1$ or an $g-$fold connected sum $({\mathbb T}^2)^{\# g}$.
\footnote{
The integer $g$ is called the {\it genus} of the surface $X$
and determines it and its fundamental group up to homeomorphism; we assume that $g\geq 1$.}
If two-manifold $X$ is not orientable, then it is the real projective plane ${\mathbb R}P^2$ or $g-$fold connected sum thereof, $({\mathbb R}P^2)^{\# g}$. Surfaces of $g\geq 2$ are covered by the genus-two double-torus, which is hyperbolic, since it is the quotient of an octagon in hyperbolic disc.
The integral homologies of surfaces of genus $g$ are
presented in Table \ref{Table1}.
\begin{table} \label{Table1}
\begin{center}
\begin{tabular}
{l l l l}
Table \ref{Table1}. Two-space integral homologies 
\\
\\
\hline
\\
Homologies \,\,\, & \,\,\, n= 0 \,\,\, & \,\,\, n= 1 & \,\,\, n = 2 \\
\\
\hline
\\
$H_n(X_{\rm oriented})$ \,\,\, & \,\,\,\,\,\,\, ${\mathbb Z}$ \,\,\,& \,\,\,\,\,\,\, ${\mathbb Z}^{\oplus 2g}$ \,\,\, & \,\,\,\,\,\,\,
${\mathbb Z}$\\
$H_n(X_{\rm non-oriented})$ \,\,\, & \,\,\,\,\,\,\, ${\mathbb Z}$ \,\,\, & \,\,\,\,\,\, ${\mathbb Z}^{\oplus g-1}
\oplus {\mathbb Z}/2$ \,\,\, & \,\,\,\,\,\,\, $0$
\\
\\
\hline
\end{tabular}
\end{center}
\end{table}
If $X$ is the orientable surface then $X= {\mathbb S}^2$
for $g=0$, and $X= {\mathbb T}^2$ for $g= 1$. For non-orientable surface
$X$ we have $X={\mathbb R}P^2$ and $X=$ the Klein bottle for
$g = 1$ and $2$, respectively. For orientable and non-orientable cases the Euler characteristic
$
\chi(X) = \sum_\ell (-1)^\ell {\rm rk}\,H_\ell(X)
$
satisfies
$
\chi(X) = 2-2g
$
and
$
\chi(X) = 2-g,
$
respectively.

\subsection{Three-geometries}
Following the presentation of \cite{Thurston,Scott}, by a geometry or a geometric structure we mean a pair $(X,\Gamma)$ that is a
manifold $X$ and a group  $\Gamma$ acting transitively 
on $X$ with compact point stabilizers (following \cite{Thurston} we also propose that the interior of every compact three-manifold has a canonical decomposition into pieces which have geometric structure). Two geometries $(X,\Gamma)$  and  $(X',\Gamma')$ are
equivalent if there is a diffeomorphism of $X$ with $X'$ which
throws the action of $\Gamma$ onto the action of $\Gamma'$. In
particular, $\Gamma$ and $\Gamma'$ must be isomorphic. 
Assume that:
\begin{itemize}
\item{} The manifold $X$ is simply connected. Otherwise it
will be sufficient to consider a natural geometry
$(\widetilde{X},\widetilde{\Gamma})$, $\widetilde{X}$ being the universal covering of $X$ and $\widetilde{\Gamma}$ denoting the group
of all diffeomorphisms of $\widetilde{X}$  which are lifts of
elements of $\Gamma$.
\item{} The geometry admits a compact quotient. In another words,
there exists a subgroup $\widehat{\Gamma}$ of $\Gamma$ which acts on $X$ as a covering group and has compact quotient.
\item{} The group  $\Gamma$ is maximal. Otherwise, if
$\Gamma\subset\Gamma'$ then any geometry $(X,\Gamma)$ would be
the geometry $(X,\Gamma')$ at the same time.
\end{itemize}
Due to Thurston's conjecture there are eight model spaces in three dimensions: 
\begin{equation*}
{X}= G/K =\left\{ \begin{array}{ll} 
{\mathbb R}^3\, {\rm (Euclidean \,\,\,\, space)}\,,\,\,\,\,\,
{\mathbb S}^3\, {\rm (spherical \,\,\, space)}\,, \,\,\,\,\,
{\mathbb H}^3 \,{\rm (hyperbolic \,\,\, space)}\,, 
\\
{\mathbb H}^2\times{\mathbb R}\,,\,\,\,\,\,
{\mathbb S}^2\times {\mathbb R}\,,\,\,\,\,\,
\widetilde{{\bf S}{\bf L}(2,{\mathbb R})}\,,\,\,\,\,\,
{\mathbb N}il^3\,,\,\,\,\,\,
{\mathbb S}ol^3\,
\end{array} \right\}
\end{equation*}
\begin{remark}
This conjecture follows from considering the identity component of the isotropy group $\Gamma_x$ of $X$ through a point $x$. $\Gamma_x$ is a compact, connected Lie group, and 
there are three cases: $\Gamma_x = {\bf SO}(3),\, {\bf SO}(2)$
and $\{ 1 \}$. 
\begin{enumerate}[${\cR} 1.$]
\item{} $\Gamma_x = {\bf SO}(3)$. In this case the space $X$
is a space of constant curvature: ${\mathbb R}^3,\, {\mathbb S}^3$ (modelled on ${\mathbb R}^3$) or ${\mathbb H}^3$ (which can be modelled on the half-space ${\mathbb R}^2\times {\mathbb R}^{+}$). 
\item{} $\Gamma_x = {\bf SO}(2)$. In this case there is a one-dimensional subspace of $T_xX$ that invariant under $\Gamma_x$, which has a complementary plane field ${\cP}_x$.
If plane field ${\cP}_x$ is integrable, then $X$ is a product
${\mathbb R}\times {\mathbb S}^2$ or ${\mathbb R}\times {\mathbb H}^2$. If plane field ${\cP}_x$ is non-integrable, then $X$ is a non-trivial fiber bundle with fiber $S^1$:
$S^1 \hookrightarrow X \twoheadrightarrow \Sigma_{g\geq 2}$
($\widetilde{{\bf S}{\bf L}(2,{\mathbb R})}-$geometry),\, 
$\Sigma_g$ stands for a surface of genus $g$,\,
$S^1 \hookrightarrow X \twoheadrightarrow {\mathbb T}^2$
(${\mathbb N}il^3-$geometry)\, or
$S^1 \hookrightarrow X \twoheadrightarrow {\mathbb S}^2$\,
(${\mathbb S}^3-$geometry).
\item{} $\Gamma_x = \{1\}$. In this case we have  three-dimensional Lie groups: 
$\widetilde{{\bf S}{\bf L}(2,{\mathbb R})}\,,\,
{\mathbb N}il^3$\,,\, and ${\mathbb S}ol^3$\,.
\end{enumerate}
\end{remark}
The first five geometries are
familiar objects, so we briefly discuss the last three ones.
The group $\widetilde{{\bf S}{\bf L}(2,{\mathbb R})}$ is the universal covering of
${\bf S}{\bf L}(2,{\mathbb R})$, the three-dimensional Lie group of all $2\times 2$
real matrices with unit determinant.
The geometry of ${\mathbb N}il$ is the three-dimensional Lie
group of all $3 \times 3$ real upper triangular matrices with
ordinary matrix multiplication. It is also
known as the nilpotent Heisenberg group.
The geometry of ${\mathbb S}ol$ is the three-dimensional
(solvable) group. Many three-manifolds are
hyperbolic (according to a famous theorem by Thurston
\cite{Thurston}). For example, the complement of a knot in
${\mathbb S}^3$ admits an hyperbolic structure unless it is a torus or satellite knot. Moreover, after the Mostow Rigidity Theorem \cite{Mostow}, any geometric invariant of an hyperbolic
three-manifold is a topological invariant.
If a three-manifold $X$ admits a geometric 
structure then the universal cover ${\widetilde X}$ with the induced metric is isometric to one of the eight geometries above. $X$ can admit more than one geometric structure, but if $X$ is closed and admits a geometric structure then the geometric structure is unique: ${\widetilde X}$ is isomorphic to one and only one of the above geometries (for more detail see \cite{Scott}). 
The Euclidean space, being flat, does not lead to a new background in supergravity, i.e. the topological twist is trivial. 
Our special interest to hyperbolic spaces: 
It has been shown that for supergravity descriptions of branes wrapping three-dimensional cycles implies that cycles have to be a constant scalar curvature. The solution including ${\mathbb H}^3$ and its quotients by subgroup of isometry group 
${\bf PSL}(2, {\mathbb C})\equiv 
{\bf SL}(2, {\mathbb C})/\{\pm 1\}$ the reader can found in \cite{Bytsenko,Bonora}.

\subsection{Four-geometries}

The list of Thurston three-geometries has been organized in terms of the compact stabilizers $\Gamma_x$ of $x\in X$ isomorphic to
${\bf SO}(3)$, ${\bf SO}(2)$ or trivial group $\{1\}$. The analogue list of
four-geometries also can be organized (using only connected
groups of isometries) as in Table \ref{Table2}.
\begin{table}\label{Table2}
\begin{center}
\begin{tabular}
{l l}
Table \ref{Table2}. List of four-geometries
\\
\\
\hline
\\
Stabilizer-subgroup $\Gamma_x$ & Space \, $X$ \\
\\
\hline
\\
${\bf SO}(4)$ & ${\mathbb S}^4,\,\,{\mathbb R}^4,\,\, {\mathbb H}^4$ \\
${\bf U}(2)$ & ${\mathbb C} P^2,\,{\mathbb C}{\mathbb H}^2$ \\
${\bf SO}(2)\times {\bf SO}(2)$ & ${\mathbb S}^2\times {\mathbb R}^2,\,\,
{\mathbb S}^2\times
{\mathbb S}^2,\,\,{\mathbb S}^2\times {\mathbb H}^2,\,\, {\mathbb H}^2\times
{\mathbb R}^2,\,\,{\mathbb H}^2\times {\mathbb H}^2$ \\
${\bf SO}(3)$ & ${\mathbb S}^3\times {\mathbb R},\,\, {\mathbb H}^3\times
{\mathbb R}$ \\
${\bf SO}(2)$ & ${\mathbb N}il^3\times
{\mathbb R},\,\,{\widetilde{{\bf PSL}}}(2,{\mathbb R})\times {\mathbb R},\,\,{\mathbb S}ol^4$ \\
${\bf S}^1$ & $F^4$ \\
{\rm trivial} & ${\mathbb N}il^4,\,\, {\mathbb S}ol^4_{m,n}$ (including
${\mathbb S}ol^3\times {\mathbb R}), {\mathbb S}ol^4_1$ \\
\\
\hline
\end{tabular}
\end{center}
\end{table}
Here we have the four irreducible four-dimensional Riemannian symmetric spaces: sphere ${\mathbb S}^4$,
hyperbolic space ${\mathbb H}^4$, complex projective space ${\mathbb C}
P^2$ and complex hyperbolic space ${\mathbb C} {\mathbb H}^2$ (which we
may identify with the open unit ball in ${\mathbb C}^2$ with
an appropriate metric). The other cases are more specific and 
only for the sake of completeness we shall illustrate them.

The nilpotent Lie group ${\mathbb N}il^4$ can be presented as the split
extension ${\mathbb R}^3 \rtimes_U {\mathbb R}$ of
${\mathbb R}^3$ by ${\mathbb R}$, where the real 3-dimensional representation $U$ of $\mathbb R$ has the form  
$U(t)=\exp (tB)$ with
$
B=\left( \begin{array}{ccc} 0 & 1 & 0\\ 0 & 0 & 1\\ 0
& 0 & 0 \end{array} \right)\,.
$
In the same way the soluble Lie groups 
${\mathbb S}ol_{m,n}^4= {\mathbb R}^3 \rtimes _{T_{m,n}}{\mathbb R}$
forms on real 3-dimensional representations $T_{m,n}$ of ${\mathbb R}$, $T_{m,n}(t)=\exp (tC_{m,n})$, where
$
C_{m,n}=
{\rm diag}(\alpha, \beta, \gamma)
$
and $ \alpha+\beta+\gamma=0 $ for $\alpha>\beta>\gamma$. Furthermore
$e^{\alpha}$, $e^{\beta}$ and $e^{\gamma}$ are the roots of
$\lambda^3-m\lambda^2+n\lambda-1=0$, with $m,\;n$ positive integers.
If $m=n$, then $\beta=0$ and  ${\mathbb S}ol_{m,n}^4= {\mathbb S}ol^3\times {\mathbb R}
$. In general, if $C_{m,n}\propto C_{m',n'}$, then ${\mathbb S}ol_{m,n}^4\cong
{\mathbb S}ol_{m',n'}^4$. It gives infinity many classes of equivalence. 
When $m^2n^2+18=4(m^3+n^3)+27$, one has a new
geometry, ${\mathbb S}ol^4_0$, associated with the group ${\bf SO}(2) $ of isometries rotating the first two coordinates. The soluble group ${\mathbb S}ol^4_1$, is most conveniently represented as the matrix group
$
\left\{\left(\begin{array}{ccc} 1 & b & c\\ 0 &
\alpha & a\\ 0 & 0 & 1 \end{array} \right)\,
: \,\,\,\alpha, a, b, c \in {\mathbb R}, \alpha> 0 \right\}\,.
$
Finally, the geometry $F^4$ is associated with the isometry group
${\mathbb R}^2 \rtimes {\bf PSL}(2,{\mathbb R})$ and stabilizer ${\bf SO}(2)$. Here the semidirect product is taken with respect to action of the group 
${\bf PSL}(2,{\mathbb R})$ on ${\mathbb R}^2$.
The space $F^4$ is diffeomorphic to ${\mathbb R}^4$
and has alternating signs of metric. A connection of these geometries with complex and K\"ahlerian structures
(preserved by the stabilizer $\Gamma_{\sigma}$) can be found in \cite{Wall}.

\section{KK-groups and K-amenable groups}

In this section we discuss K-groups of the twisted group $C^*-$algebras, which are relevant to the definition of a twisted analogue of the Kasparov map, pairing between K-theory and cyclic cohomology theory, and which enables us to use 
K-amenability results for Eilenberg-Maclane spaces.

\subsection{KK-groups}
Given a manifold $X$, let $C(X)$ be the commutative $C^*-$algebra
(recall that a $C^*-$algebra is a Banach algebra with an
involution satisfying the relation $||aa^*||=||a||^2$) of all
continuous complex-valued functions which vanish at infinity on
$X$. The $C^*-$algebra, which categorically encodes the
topological properties of manifold $X$, plays a dual role to $X$ in the K-theory of $X$ by the Serre-Swan theorem \cite{WO}:
\begin{equation}
\widetilde{K}^\ell(X)\cong \widetilde{K}_{\ell}(C(X)),\,\,\,\,\, \ell = 0, 1\,.
\end{equation}
Here $\widetilde{K}^\ell(X)$ is the reduced topological K-theory of $X$: Taking into account that a vector bundle
over a point is just a vector space, 
$K({\rm pt})= {\mathbb Z}$,
we can introduce a reduced K-theory in which the topological space consisting of a single point has trivial cohomology,
${\widetilde{K}}({\rm pt})=0$, and also 
${\widetilde {K}}(X)=0$
for any contractible space $X$. Let us consider the collapsing and inclusion maps: 
$ p:X\rightarrow {\rm pt}\,,\,\,\iota: {\rm
pt}\hookrightarrow X $ for a fixed base point of $X$. These maps
induce an epimorphism and a monomorphism of the corresponding
K-groups: $ p^*: K({\rm pt}) = {\mathbb Z}\rightarrow K(X)\,,$ 
$\iota^*: K(X)\rightarrow K({\rm pt})= {\mathbb Z}.$ The exact
sequences of groups are:
$$
0\rightarrow {\mathbb Z} \stackrel{p^*}{\rightarrow} K(X)
\rightarrow {\widetilde{K}}(X)
\rightarrow 0\,,\,\,\,\,\,\,\,
0 \rightarrow {\widetilde {K}}(X)\rightarrow K(X)
\stackrel{\iota^*}{\rightarrow} {\mathbb Z}\,.
$$
The kernel of the map $i^*$ (or the cokernel of the map $p^*$) is
called the {\it reduced K-theory group} and is denoted by
${\widetilde {K}}(X)$, ${\widetilde {{K}}}(X) = {\rm ker} \iota^* = {\rm coker}\,p^*$. There is a fundamental
decomposition $K(X)= {\mathbb Z}\oplus \widetilde{{K}}(X)$. 
When $X$ is not compact, we can define $K^c(X)$, the K-theory with compact support. It is isomorphic to ${\widetilde{{K}}}(X)$.
Let $X$ be a $Spin^{\mathbb C}-$manifold, then
there is a Poincar\'{e} duality isomorphism \cite{Higson}:
\begin{equation}
K^{{\rm dim}\,X-\ell}(X)\cong K_{\ell}^{c}(X)\,,
\,\,\,\,\,\,\, \ell =0, 1\,,
\end{equation}
where $K_{\ell}^{c}$ denotes the dual compactly supported K-homology of $X$.

For a finite dimensional manifold $X$ there exists another
$C^*-$algebra, which is non-commutative and can be constructed
with the help of the Riemannian metric $g$. In fact, we can form
the complex Clifford algebra ${\rm Cliff}(T_xX,g_x)$, where for
each $x\in X$ the tangent space $T_xX$ of $X$ is a
finite-dimensional Euclidean space with inner product $g_x$. This algebra has a canonical structure as a finite-dimensional
${\mathbb Z}_2-$graded $C^*-$algebra. The family of $C^*-$algebras 
$\{{\rm Cliff}(T_xX, g_x\}_{x\in X}$ forms a
${\mathbb Z}_2-$graded $C^*-$algebra vector bundle ${\rm
Cliff}(T_xX)\rightarrow X$, called {\it the Clifford algebra bundle} of $X$ \cite{Atiyah64}. Let us define 
${\frak C}(X) = C(X, {\rm Cliff}(T_xX))$ to be the $C^*-$algebra of continuous sections of the
Clifford algebra bundle of $X$ vanishing at infinity. 
If the manifold $X$ is even-dimensional and has a $Spin^{\mathbb C}-$structure then this $C^*-$algebra is Morita equivalent to $C(X)$. However, in general, 
${\frak C}(X)$ is Morita equivalent to $C(T_xX)$. 
Because of the Morita equivalence of K-theory, it follows that
$
K_{\ell}({\frak C}(X))\cong
K_{\ell}(C(X))\cong K^{\ell}(X)\,,\,\ell =0, 1\,.
$ 
But for odd-dimensional and spin manifold $X$ this relation is more complicated.

Recall that the definition of K-homology
involves classifying extensions of the algebra of continuous
functions $C(X)$ on the manifold $X$ by the algebra of compact
operators up to unitary equivalence \cite{Brown}. The set of
homotopy classes of operators defines the K-homology group
$K_0(X)$, and the duality with K-theory is provided by the natural
bilinear pairing $ ([E]\,,\,[{\mathfrak D}])\mapsto {\rm Index}\,
{\mathfrak D}_{E}\in {\mathbb Z}\,, $ where $[E]\in K(X)$ and
${\mathfrak D}_E$ denotes the action of the Fredholm operator
${\mathfrak D}$ on the Hilbert space ${\mathfrak H}= L^2(U(X,E))$
of square-integrable sections of the vector bundle $E\rightarrow
X$ as ${\mathfrak D}: U(X,E)\rightarrow U(X,E)$. It assumes that the KK-pairing may be the most natural framework in this context. The group $KK(A,B)$ is a bivariant version of 
K-theory and it depends on a pair of graded algebras $A$ and $B$.
\footnote{
The reader can find the application of KK-group theory to the classification of branes, for example, in \cite{Periwal,Brodzki}.
}
\begin{definition}
Let $A$ and $B$ be $C^*-$algebras. 
\begin{enumerate}[${\cD} 1.$]
\item{} A pair $({\mathcal E},
\pi)$ will be called an $(A,B)-${\it bimodule} if ${\mathcal E}$ is a ${\mathbb Z}/2{\mathbb Z}$ graded Hilbert $B-$module on
which algebra $A$ acting by means of \, $*-$homomorphism
$\pi$: 
$
A\rightarrow {\mathfrak L}({\mathcal E}) = 
{\rm End}^*({\mathcal E})\,,
$
where $\forall a\in A$ an operator $\pi (a)$ being of degree 0, $\pi (A)\subset {\mathfrak L}({\mathcal E})^{(0)}$.
Let $E(A,B)$ be a triple $({\mathcal E}, \pi , F)$,
where $({\mathcal E}, \pi)$ is a $A,B-$bimodule, 
$F\in {\mathfrak L}({\mathcal E})$ is a homogeneous operator of degree 1, and $\forall a \in A$:
$
\pi (a)(F^2-1)\in {C}({\mathcal E})\,,\,
[\pi (a), F] \in {C}({\mathcal E})\,,
$
where ${C}({\mathcal E})$ is the algebra of compact operators.
\item{} A triple $({\mathcal E}, \pi , F)$ will be called degenerate if
$\forall a \in A$: 
$
\pi (a)(F^2-1) = 0\,,\, 
[\pi (a), F] = 0\,.
$
\item{} Let $D(A, B)$ be a set of degenerated triples. 
An element 
$E(A, B[0,1])$, where $B[0,1]$ is an algebra of continuous functions in $B$ on the interval $[0,1]$, will be called a homotopy in $E(A,B)$. 
\end{enumerate}
\end{definition}
Let us assign a direct sum in $E(A,B)$:
\begin{equation}
({\mathcal E}, \pi ,
F)\oplus({\mathcal E}^{'}, \pi^{'} , F^{'}) = ({\mathcal
E}\oplus{\mathcal E}^{'},\pi\oplus\pi^{'},F\oplus F^{'})\,.
\end{equation}
The homotopy classes of $E(A,B)$ together with this sum defines
the Abelian group $KK(A,B)$. It is clear that any degenerate triplet is homotopy equivalent to $(0,0,0)$, and the inverse element to $({\mathcal E}, \pi, F)$ is equal to 
$(-{\mathcal E}, \pi,-F)$, where $-{\mathcal E}$ means that grading on ${\mathcal E}$ has to be inverted.
$f: A_1\rightarrow A_2$ transfers
$(A_2, B)-$modules into $(A_1, B)-$modules, and \cite{Kasparov1}
\begin{equation}
f^*: E(A_2,B)\rightarrow E(A_1,B)\,,\,\,\,\,\,\,\,
({\mathcal E},\pi,F)\mapsto ({\mathcal E},\pi^{\circ}f,F)\,,
\end{equation}
On the other hand a $*-$homomorphism $g:
B_1\rightarrow B_2$ induces a homomorphism
\begin{equation}
g_*: \, E(A,B_1)\rightarrow E(A,B_2)\,,\,\,\,\,\,\,\,
({\mathcal E},\pi,f) \mapsto
({\mathcal E}\otimes_gB_2,\pi\otimes 1,F\otimes 1)\,,
\end{equation}
where
\begin{equation}
\pi\otimes 1: A\rightarrow {\mathcal L}({\mathcal E}\otimes_gB_2)\,,
\,\,\,\,\,\,\,
(\pi\otimes 1)(a)(e\otimes b)=\pi(a)e\otimes b\,.
\end{equation}
\begin{theorem}
The groups
$KK(A,B)$ define an homotopy invariant bifunctor from the category of
separable $C^*-$algebras into the category of Abelian groups.
Abelian groups $KK(A,B)$ depend contravariantly on the algebra
$A$ and covariantly on the algebra $B$, in addition 
$KK({\mathbb C}, B)= K_0(B)$.
\footnote{
The interesting facts to us are the following relations:
$
K\!K^{*}({A:=\mathbb C}, B) = K_{*}(B), 
K\!K^{*}(A, B:={\mathbb C}) = K^{*}(A).
$
}
\end{theorem}
\begin{definition}
Let  $1_A \in KK(A,A)$ ($KK(A,A)$ is a ring with unit) denotes the triple class $(A, \iota_A,0)$ where $A^{(1)}=A,\, A^{(0)}=0$, and
$\iota_A: A\rightarrow {C}(A)\subset {\mathfrak L}(A)$,
$\iota_A(a)b = ab,\, a,b\in A$.
Let us define also the map
\begin{enumerate}[${\cD} 1.$]
\item{} $\tau_D : \,\,
KK(A,B)\otimes KK(A\otimes D, B\otimes D)\,.$
\item{} $\tau_D\,({\rm class}\,\,({\mathcal E}, \pi, F)) =
{\rm class}\,\, ({\mathcal E}\otimes D, \pi \otimes 1_D, F \otimes 1)\,.$
\end{enumerate}
\end{definition}
\begin{theorem} Kasparov's pairing, defined by
\begin{equation}
KK(A,D)\times KK(D,B)\longrightarrow KK(A,B)
\end{equation}
and denoted $(x,y)\mapsto x\otimes_Dy$,
satisfies the following properties:
\begin{enumerate}[${\cT} 1.$]
\item{} It depends covariantly on
the algebra $B$ and contravariantly on the algebra $A$.
\item{} If $f: D\rightarrow E$ is a $*-$homomorphism, then
$f_*(x)\otimes_Ey =x\otimes_Df^*(y)$,\,
$x\in KK(A,D),\, y\in KK(E, B)$.
\item{} Associative property:
$(z\otimes_Dy)\otimes_Ez=x\otimes_D(y\otimes)_Ez$,\,
$\forall x\in KK(A,D),\, y\in KK(D,E),\, z\in KK(E,B)$.
\item{} $x\otimes_B1_B=1_A\otimes x =x,\, \forall x\in KK(A,B)$.
\item{} $\tau_E(x\otimes_{B} y)=\tau_E(x)\otimes_{B\otimes E}\tau_E(y)\,,
\,\, \forall x\in KK(A, B),\, \forall y \in KK(B,D)$.
\end{enumerate}
\end{theorem}
Suppose that for two algebras, $A$ and $B$, there are elements
$\alpha\in K\!K(A\otimes B,{\mathbb C})$, $\beta\in K\!K({\mathbb
C},A\otimes B)$, with the property that $\beta\otimes_{A}\alpha =
1_{B} \in\ K\!K(B,B)$,  $\beta\otimes_{B}\alpha = 1_{A} \in\
K\!K(A,A).$ Then we say that we have KK-duality isomorphisms between
the K-theory (K-homology) of the algebra $A$ and the K-homology
(K-theory) of the algebra $B$
\begin{equation}
K_{*}(A) \cong K^{*}(B)\, ,\,\,\,\,\,\,\,
K^{*}(A) \cong K_{*}(B)\,.
\end{equation}
In fact, the algebras $A$ and $B$ are Poincar\'e dual
\cite{Connes94}, but generally speaking these algebras are not
KK-equivalent.

\subsection{K-amenable groups}
We now review the concept of K-amenable
groups \cite{Carey}. Let $G$ be a connected Lie group and
${K}$ a maximal compact subgroup. We also assume that
${\rm dim}\, (G/K)$ is even and $G/K$ admits
a $G-$invariant ${Spin}^{\mathbb C}-$structure. The
$G-$invariant Dirac operator ${\mathfrak D :=
\gamma^{\mu}\partial_{\mu}}$ on $G/K$ is a first order
self-adjoint, elliptic differential operator acting on $L^2-$
sections of the ${\mathbb Z}_2-$graded homogeneous bundle of
spinors $\mathfrak S$. Let us consider a $0^{\rm th}$ order
pseudo-differential operator 
${\cO} = {\mathfrak D(1+{\mathfrak D}^2)^{-\frac 12}}$ acting on $H=L^2(G/K, {\mathfrak S})$. $C(G/K)$ acts on $H$ by
multiplication of operators. $G$ acts on $C(G/K)$ and
on $H$ by left translation, and ${\cO}$ is $G-$invariant.
Then, the set $({\cO}, H, X)$ defines a canonical Dirac
element $\alpha_G = KK_G(C(G/K), {\mathbb C})$.
\begin{theorem} [{\it G. Kasparov} \cite{Kasparov95}]
There is a canonical Mishchenko element
\begin{equation}
\alpha_G \in KK_G(C(G/K), {\mathbb C})
\end{equation}
such that the following intersection products occur:
\begin{enumerate}[${\cT} 1.$]
\item{} $\alpha_G\otimes_{\mathbb C}\beta_G = 1_{C(G/
K)} \in KK_G(C(G/K),C(G/K))$\,. 
\item{}
$\beta_G\otimes_{C(G/K)}\alpha_G=\gamma_G=
KK_G({\mathbb C}, {\mathbb C})$ where $\gamma_G$ is an element in
$KK_G({\mathbb C}, {\mathbb C})$.
\end{enumerate}
\end{theorem}
All solvable groups are amenable, while any non-compact semisimple Lie group is non-amenable.
For a semisimple Lie group $G$ or for $G={\mathbb R}^n$, a
construction of the Mishchenko element $\beta_G$ can be found in
\cite{Carey}. We now come to the basic theorem and definition:
\begin{theorem} [{\it G. Kasparov} \cite{Kasparov95}]
If group $G$ is amenable, then $\gamma_G=1$.
\end{theorem}
\begin{definition} A Lie group $G$ is said to be
K-amenable if $\gamma_G$=1.
\end{definition}
\begin{proposition} \label{Pr11}
The following statements hold:
\begin{enumerate}[${\cP} 1.$]
\item{} Any solvable Lie group and any amenable Lie group is K-amenable. 
\item{} The non-amenable groups ${\mathbf{SO}}_0(n,1)$
are K-amenable Lie groups {\rm \cite{Kasparov}}.
\item{} The groups ${\mathbf{SU}}(n,1)$ are K-amenable Lie groups {\rm \cite{Julg}}.
\item{} The class of K-amenable groups is closed under the operations of taking subgroups, under free and direct products {\rm \cite{Cuntz}}.
\end{enumerate}
\end{proposition}

\section{Aspects of K-group theory}

\subsection{K-theory of twisted group $C^*-$algebras}

{\bf Twisted crossed products of $C^*-$algebras
\footnote{
It is often useful to introduce $C^*-$algebras $A$ as involute
Banach algebras for which the following equalities hold
$(xy)^*=y^*x^*,\, ||x^*x|| = ||x||^2$ for $x, y \in A$. 
A unique norm is given for any $x$ by $||x||$ = (spectral radius of $(x^*x)^{1/2}$)}}.
Let us consider a general family of twisted actions of locally compact groups on $C^*-$algebras, and the corresponding twisted crossed product $C^*-$algebras.
We start with the definition of a twisted action of a locally compact group $G$ on $C^*-$algebra $A$ (see for detail Ref. \cite{Packer}). Let ${\rm Aut} A$
and $UM(A)$ denote its automorphism group and the group of unitary elements in its multiplier algebra $M(A)$.
A twisted action of $G$ on $A$ is a pair ${\alpha, u}$ of Borel maps
$\alpha: G\rightarrow {\rm Aut}A\,,\, u: G\times G\rightarrow UM(A)$ satisfying 
$
\alpha_s\circ \alpha_t = {\rm Ad} u(s, t)\circ \alpha_{st}\,, 
\alpha_r(u(s, t))u(r, st) = u(r, s)u(rs, t)\,.
$
These twisted actions and a twisted Banach $*-$algebra
$L^1(A, G, \alpha, u)$ have been introduced in \cite{Busby}. 
The quadruple $(A, G,\alpha,u)$ can be refered as a 
{\it (separable) twisted dynamical system}:
\begin{enumerate}[${\bullet} $]
\item{} The covariant theory of the system $(A, G,\alpha,u)$
can be realized on Hilbert space. The corresponding reduced crossed product can be defined as the $C^*-$algebra generated by the regular representation \cite{Packer}. In \cite{Quigg} a duality theorem has been proved for that 
$C^*-$algebra.
\item{} The twisted cross product $A\times_{\alpha, u}G$
was defined as a $C^*-$algebra whose representation theory is the same as the covariant representation theory of 
$(A,G,\alpha,u)$ on Hilbert space \cite{Packer}. A cross product by coactions of possibly non-amenable groups also has been considered in \cite{Raeburn}.
\item{} Suppose that $(\alpha,u)$ is a twisted action of an amenable group $G$ on a $C^*-$algebra $A$ which is the algebra of sections of a $C^*-$bundle $\cE$ over $X$; each $\alpha$ leaves all ideals $I_x=\{a\in A: a(x)=0\}$ invariant. It has been shown \cite{Packer2} that $A\times_{\alpha, u}G$ is the algebra of sections of a $C^*-$bundle over $X$ with fibers of the form $(A/I_x)\times_{\alpha(x), u(x)}G$. This result has been proved by using the stabilization trick of \cite{Packer}.

{\it Stabilisation trick}: The twisted cross product algebra $A\rtimes_\sigma \Gamma$ is stably equivalent to the cross product 
$(A\otimes \cK )\rtimes \Gamma$, where $\cK$ denotes the compact operators (we refer the reader to the article \cite{Packer} for details).
\end{enumerate}

{\bf Twisted K-groups}. Let $\Gamma$ be a discrete cocompact subgroup of a solvable simply-connected Lie group $G$. 
It has been shown in \cite{Rosenberg} that 
\begin{equation}
K_*(C^*(\Gamma))\cong K^{*+{\rm dim} G}(G/\Gamma)\,.
\end{equation}
For a multiplier $\sigma$ on $\Gamma$ (a cocycle $\sigma\in Z^2(\Gamma, {\bf U}(1))$) the K-theory of the twisted group algebra $C^*(\Gamma,\sigma)$ is that of a continuous-trace $C^*-$algebra $B_\sigma$ with spectrum $G/\Gamma$, i.e. the twisted K-theory $K^*(G/\Gamma, \delta(B_\sigma))$ \cite{Packer2}. The Disxmier-Douady class $\delta(B_\sigma)$ can be identified as the image of $\sigma$ under a homomorphism \cite{Wassermann,Raeburn2}:
\begin{equation}
\delta: H^2(\Gamma, {\bf U}(1)) \longrightarrow H^3(G/\Gamma, {\mathbb Z})\,,
\end{equation}
it depends only on the homotopy class of $\sigma$ in $H^2(\Gamma, {\bf U}(1))$. This result has been extended in \cite{Packer2} to describe the K-theory of the twisted transformation group algebras $C_0(X)\times_{\tau,\omega}\Gamma$, where $X$ is a $\Gamma-$space and $\omega\in Z^2(\Gamma, C(X, {\bf U}(1)))$.
Thus $K_*(C_0(X)\times_{\tau, \omega}\Gamma)$ is isomorphic to a twisted K-group $K^*((G\times X)/\Gamma, \delta(\omega))$
of the orbit space $(G\times X)/\Gamma$ for the diagonal action, and identify the twist 
$\delta(\omega)\in H^3((G\times X)/\Gamma, {\mathbb Z})$.
Further generalisation can be obtained for describing the twisted transformation group algebra 
$C_0(Y)\times_{\tau, \omega}G$ associated with a locally compact group $G$, a principal $G-$bundle $Y$ and cocycle $\omega\in Z^2(G, C(Y, {\bf U}(1)))$ \cite{Packer2}.

Let $\Gamma\subset G$ be a lattice in $G$ and $A$ be an algebra
admitting an automorphic action of $\Gamma$.
The cross product algebra $[A\otimes C_0(G/K)]\rtimes \Gamma$
is Morita equivalent to the algebra of continuous sections
vanishing at infinity $C_0(\Gamma\backslash G/K, E)$.
Here $E\rightarrow \Gamma\backslash G/K$ is the flat $A-$bundle defined as the quotient
\begin{equation}
E = (A\times G/K)/\Gamma \longrightarrow \Gamma\backslash G/K\,,
\end{equation}
and we consider the diagonal action of $\Gamma$ on $A\times G/K$.
\begin{theorem}[{\it G. Kasparov}
\cite{Kasparov95}]\label{Th1}
Let $G$ be a K-amenable, then
$(A\rtimes\Gamma)\otimes C_0(G/K)$ and $[A\otimes C_0(G/K)]\rtimes
\Gamma$
have the same K-theory.
\end{theorem}
\begin{corollary} \label{CC}
%[{\it A. L. Carey, K. C. Hannabuss, V. Mathai %and P. McCann} \cite{Carey}]
Let $G$ be a K-amenable, then $(A\rtimes\Gamma)\otimes C_0(G/K)$ and $C_0(\Gamma\backslash G/K, E)$ have the same K-theory. It means that for $\ell =0,1$ one has
\begin{equation}
K_\ell(C_0(\Gamma\backslash G/K, E)) \cong K_{\ell+ \dim(G/K)}
(A\rtimes\Gamma)\,.
\end{equation}
\end{corollary}
We follow the lines of the article \cite{Carey} in the formulation and in the proof of the main theorem which generalizes theorems of \cite{Packer} and \cite{Packer2}.
\begin{theorem} [{\it A. L. Carey, K. C. Hannabuss, V. Mathai and P. McCann} \cite{Carey}]\label{main}
Let $\Gamma$ be a lattice in a K-amenable Lie group $G$ and 
$K$ be a maximal compact subgroup of $G$. Then
\begin{equation}
K_*(C^*(\Gamma,\sigma)) \cong K^{* + \dim(G/K)}(\Gamma\backslash G/K,
\delta(B_\sigma))\,,
\label{K-am}
\end{equation}
where $\sigma \in H^2(\Gamma, U(1))$ is any multiplier on $\Gamma$,
$K^{*}(\Gamma\backslash G/K,
\delta(B_\sigma))$ is the twisted K-theory of a continuous trace $C^*-$algebra
$B_\sigma$ with spectrum $\Gamma\backslash G/K$, and $\delta(B_\sigma)$ denotes the
Dixmier-Douady invariant of $B_\sigma$.
\end{theorem}
{\it Proof.} First suppose that $A={\mathbb C}$ and $\Gamma$ acting trivially on ${\mathbb C}$. Because of the Corollary \ref{CC} when $\gamma_G=1$ one gets
\begin{eqnarray}
&&
\underbrace{
\left\{
\begin{array}{ll} 
(A\rtimes \Gamma)\otimes C_0(G/K)
\\
C_0(\Gamma\backslash G/K, E)
\end{array} 
\right\}}_
{{{\rm have}\,\,\,{\rm the}\,\,\,{\rm same}\,\,\,
{\rm K-theory}}}
=
\underbrace{\left\{
\begin{array}{ll} 
({\mathbb C}\rtimes \Gamma)\otimes C_0(G/K)
\\
C_0(\Gamma\backslash G/K, E)
\end{array} \right\}}_
{{{\rm have}\,\,\,{\rm the}\,\,\,{\rm same}\,\,\,{\rm K-theory}}}
\stackrel{{\rm Theorem}\, \ref{Th1}}{\Longleftarrow\Longrightarrow}
\underbrace{\left\{
\begin{array}{ll} 
[{\mathbb C}\otimes C_0(G/K)]\rtimes \Gamma
\\
C_0(G/K, E)
\end{array} \right\}}_
{{{\rm have}\,\,\,{\rm the}\,\,\,{\rm same}\,\,\,{\rm K-theory}}}
\nonumber \\
&&
\Longrightarrow C^*(\Gamma)\,\,\,
{\rm and}\,\,\, C_0(\Gamma\backslash G/K)\,\,\,{\rm have}\,\,\,{\rm the}
\,\,\,{\rm same}\,\,\, {\rm K-theory}\,.
\nonumber
\end{eqnarray} 
Suppose $\sigma\in H^2(\Gamma, {\bf U}(1))$. If $G$ is K-amenable, then using Corollary \ref{CC} one sees
\begin{eqnarray}
&& 
\underbrace{
\left\{
\begin{array}{ll} 
(A\rtimes \Gamma)\otimes C_0(G/K)
\\
C_0(\Gamma\backslash G/K, E)
\end{array} 
\right\}
}_
{{\rm have}\,\,\,{\rm the}\,\,\,{\rm same}\,\,\,{\rm K-theory}}
\Longrightarrow
\underbrace{
\left\{
\begin{array}{ll} 
(A\rtimes_\sigma \Gamma)\otimes C_0(G/K)
\\
C_0(\Gamma\backslash G/K, E_\sigma)
\end{array} 
\right\}
}_
{{\rm have}\,\,\,{\rm the}\,\,\,{\rm same}\,\,\,{\rm K-theory}}\,,
\nonumber \\
&&
\,\,\,\,\,
{\rm where}\,\,\, E_\sigma= (A\otimes{\cK}\times G/K)/\Gamma\longrightarrow \Gamma\backslash G/K\,.
\nonumber
\end{eqnarray} 
Since by definition the twisted K-theory $K^*(\Gamma\backslash G/K, \delta(B_\sigma))$ is the K-theory of the continuous trace $C^*-$algebra 
$B_\sigma = C_0(\Gamma\backslash G/K, E_\sigma)$ with spectrum
$\Gamma\backslash G/K$. Thus Eq. (\ref{K-am}) follows.
$\blacksquare$

\subsection{The three-dimensional case}
One of the main result of \cite{Carey} says that for lattices in K-amenable Lie groups the reduced and unreduced twisted group $C^*-$algebras have canonically isomorphic K-theory.
If $\sigma\in H^2(\Gamma, {\bf U}(1))$ is a multiplier on $\Gamma$ and $\Gamma$ is a lattice in a K-amenable Lie group, then the canonical morphism 
$C^*(\Gamma, \sigma)\rightarrow 
C_r^*(\Gamma, \sigma)$ induces an isomorphism
\begin{equation}
K_*(C^*(\Gamma, \sigma))\cong K_*(C_r^*(\Gamma, \sigma))\,.
\end{equation}
\begin{corollary} \label{Cor1}
Suppose $G$ is a connected Lie group and $K$ a maximal compact subgroup such that ${\rm dim}(G/K) = 3$. Let $\Gamma$ be an uniform lattice in $G$ and $\sigma\in H^2(\Gamma, U(1))$ be any multiplier on $\Gamma$. Suppose also that $G$ is K-amenable, then for $\ell =0,1\,$ (mod 2)
\begin{equation}
K_\ell(C^*_r(\Gamma,\sigma)) \cong K_\ell(C^*_r(\Gamma)) \cong
K^{\ell+1}(\Gamma\backslash G/K)\,.
\label{K11}
\end{equation}
\end{corollary}
Indeed, by the main Theorem \ref{main}, 
$
K_\ell(C^*_r(\Gamma)) \cong K^{\ell+\dim(G/K)}(\Gamma\backslash G/K),
$
for $\ell\,=0,1$\, (mod\, 2).
Because of the Packer-Raeburn stabilization trick, $C_r^*(\Gamma,\sigma)$ is Morita
equivalent to $C\rtimes\Gamma$, and since $G$ is K-amenable,
$C\rtimes\Gamma\otimes C_0(G/K)$ is Morita equivalent to
$B_\sigma = C(\Gamma\backslash G/K, E_\sigma)$. Here as before $E_\sigma$ is a locally trivial bundle of $C^*-$algebras over $\Gamma\backslash G/K$ with fibre $\cK$. 
For the Dixmier-Douady invariant one has
$
\delta(B_\sigma) = \delta(\sigma) \in 
H^3(\Gamma\backslash G/K, \mathbb{Z}) 
\cong H^3(\Gamma,\mathbb{Z})\,.
$
If $\Gamma\backslash G/K$ is not orientable, then $H^3(\Gamma\backslash G/K, {\mathbb Z}) = \{0\}$. Therefore
$\delta(B_\sigma)=0$ (see Table \ref{TableK}) and Eq. (\ref{K11}) holds.
\begin{table}\label{TableK}
\begin{center}
\begin{tabular}
{llll}
Table \ref{TableK}. %$ K^* (\Gamma\backslash G/K, \delta(B_\sigma))$ 
\\
\\
\hline
\\
$K^*(\delta)\equiv K^*(\Gamma\backslash G/K, \delta(B_\sigma))$  & 
{} & {} \\
($G$ is K-amenable) &  
$\delta (B_\sigma)$ & 
$H^3(\Gamma\backslash G/K, {\mathbb Z})$ \\
\\
\hline
\\
non twisted group ${K}^*(0)$ & {} & {} \\
($\Gamma\backslash G/K$ is orientable) &
0 \,($B_\sigma$ is Morita equivalent 
to $C(\Gamma\backslash G/K)$ &
\,\,\,\,\,\,\,\,\, ${\mathbb Z}$ 
\\
\\
non twisted group ${K}^*(0)$ & {} & {} \\
($\Gamma\backslash G/K$ is not orientable) &
0 \, ($B_\sigma$ is Morita equivalent 
to $C(\Gamma\backslash G/K)$ &  
\,\,\,\,\,\,\,\,\, 0
\\
\\
twisted group ${K}^*(\delta)$ & $\neq 0$
(${K}^*(\delta)$ 
can be not isomorphic to ${K}^*(0)$, 
& {} \\
{} & but there is isomorphism for $\delta = 0$ 
\cite{Packer}) & {} 
\\
\\
\hline
\end{tabular}
\end{center}
%\caption{}
%\label{tab1}
\end{table}
On the contrary, when $\Gamma\backslash G/K$ is orientable, 
$\delta(\sigma)=0$ for all $\sigma\in H^2(\Gamma,U(1))$ \cite{Carey}, and 
$B_\sigma$ is Morita equivalent to $C(\Gamma\backslash G/K)$
(see Table \ref{TableK}). In this case again we have Eq. (\ref{K11}).
\begin{corollary} 
[{\it A. L. Carey, K. C. Hannabuss, V. Mathai and P. McCann}
\cite{Carey}]\label{Cor2}
Let \, $M \,=$
\\
$K(\Gamma,1)$ be an Eilenberg-MacLane space which is a connected locally-symmetric, compact,
three-dimensional manifold. Let $\sigma\in H^2(\Gamma, U(1))$ be any multiplier on $\Gamma$, then one has
\begin{equation}
K_\ell(C^*_r(\Gamma, \sigma)) \cong K_\ell(C^*_r(\Gamma)) \cong K^{\ell+1}(M), \quad \ell = 0,1\,.
\label{K-ei}
\end{equation}
\end{corollary}
Indeed, $K(\Gamma, 1)$ is locally symmetric (see the Appendix for the necessary information on Eilenberg-MacLane spaces), and therefore it is of the form $\Gamma\backslash G/K$, where 
$G$ is a connected Lie group, $K$ is a maximal compact subgroup such that $\dim(G/K) = 3$ and $\Gamma\subset G$ is an uniform lattice in $G$. According to Thurston's list of three-geometries or locally homogeneous spaces, one has
\begin{itemize}
\item{}  $G = \mathbb{R}^3 \rtimes {\bf SO}(3),\ G/K = \mathbb{R}^3$ (flat).
\item{}  $G={\bf SO}_0(3,1),\ G/K=\mathbb{H}^3$ (hyperbolic; compact example: non-trivial $\Sigma_{g\geq 2}-$bundle over $S^1$.
\item{}  $G={\bf SO}_0(2,1)\rtimes\mathbb{R},\ G/K = \mathbb{H}^2\times\mathbb{R}$ (hyperbolic; compact example: 
trivial $\Sigma_{g\geq 2}-$bundle over $S^1$, i.e. $\Sigma_g\times S^1$).
\item{}  $G= {\mathbb N}il^3 = {\mathbb Z}\rtimes {\mathbb R}^2$\, (central, non-split extension),\ $G/K = {\mathbb N}il^3$ (flat; compact example: ${\mathbb T}^2-$bundle over $S^1$ via mapping torus).
\item{}  $G={\mathbb S}ol^3 = {\mathbb R}^2 \rtimes {\mathbb R}$ (split extension),\ $G/K={\mathbb S}ol^3$ (flat; compact example: ${\mathbb T}^2-$bundle over $S^1$ via mapping torus).
\item{}  $G=\widetilde{{\bf SO}_0(2,1)}\rtimes\mathbb{R},\, 
G/K=\widetilde{{\bf SO}_0(2,1)}$ (hyperbolic; compact example: $S^1-$bundle over $F_{g\geq 2}$).
\end{itemize}
For all of these three-manifolds $\gamma_G=1$. More two locally homogeneous spaces in Thurston's list are not locally symmetric. We can apply Corollary \ref{Cor1} in order to deduce Corollary \ref{Cor2}.

\subsection{The four-dimensional case}
Using the mathematical tools exposed so far, we are now able to formulate new results concerning K-amenability in the four-dimensional case.
\begin{corollary}\label{Cor3}
Lie groups $G$ associated with spaces $G/K = {\mathbb R}^4, 
{\mathbb H}^4, {\mathbb C}P^2, {\mathbb C}{\mathbb H}^2,
{\mathbb H}^2\times {\mathbb R}^2\,, 
{\mathbb H}^2\times {\mathbb H}^2\,,{\mathbb H}^3\times {\mathbb R}\,, {\mathbb N}il^3\times {\mathbb R}\,,
{\mathbb S}ol^4\,,{\mathbb N}il^4\,,{\mathbb S}ol^4_{m,n}$ \,(including
${\mathbb S}ol^3\times {\mathbb R})\,, 
{\mathbb S}ol^4_1\,, 
\widetilde{{\bf PSL}(2,{\mathbb R})}\times {\mathbb R}\,,
F^4\ $ enumerated in Table \ref{Table2} are K-amenable.
\end{corollary}
{\it Proof}. We need to proof that $\gamma_G =1$. According to list of four-geometries, Table \ref{Table2}, one has the following result:
\begin{enumerate}[${\cC} 1.$]
\item{}  $G = \mathbb{R}^4 \rtimes {\bf SO}(4),\ G/K = \mathbb{R}^4,\ \gamma_G=1$
since $\mathbb{R}^4$ and ${\bf SO}(4)$ are amenable, and 
so is their semidirect product (Proposision \ref{Pr11}).
\item{}  $G={\bf SO}_0(4,1),\ G/K=\mathbb{H}^4,\ \gamma_G = 1$ by Kasparov's theorem (Proposition \ref{Pr11}).
\item{}  $G= {\bf SU}(3),\ G/K = {\mathbb C}P^2 \simeq {\bf U}(3)/({\bf U}(1)\times {\bf U}(2))\simeq {\bf SU}(3)/{\bf S}({\bf U}(1)\times {\bf U}(2)),
\gamma_G = 1$ (Proposition \ref{Pr11}).
\item{}  $G= G/K = {\mathbb C}{\mathbb H}^2$\,,\,\, 
the geometry ${\mathbb C}{\mathbb H}^2$ is a K\"ahlerian symmetric space and certainly carries a complex structure,
$\gamma_G = 1$.
\item{}  $G/K = {\mathbb H}^2\times {\mathbb R}^2\,, \,
{\mathbb H}^2\times {\mathbb H}^2\,,\,{\mathbb H}^3\times {\mathbb R}\,,\,\,
\gamma_G = 1$ since these groups are free products of K-amenable groups (Proposition \ref{Pr11}).
\item{}  $G/K = {\mathbb N}il^3\times {\mathbb R}\,,\,
{\mathbb S}ol^4\,,\,{\mathbb N}il^4\,,\,\, {\mathbb S}ol^4_{m,n}$ \,(including
${\mathbb S}ol^3\times {\mathbb R})\,,\, {\mathbb S}ol^4_1\,\,\, 
\gamma_G = 1$ since nilpotent and solvable groups are K-amenable groups and so is their semidirect product (Proposition \ref{Pr11}).
\item{}  $G/K = \widetilde{{\bf PSL}(2,{\mathbb R})}\times {\mathbb R}$\,,\,\, group $\widetilde{{\bf PSL}(2,{\mathbb R})}$ is diffeomorphic to ${\mathbb R}^3$, \, 
$\gamma_G = 1$ since $X$ is free products of K-amenable groups (Proposition \ref{Pr11}).
\item{}  $G/K = F^4$\,. One can choose a subgroup $\cG_X$ of $X$ which admits $X$ as a principal homogeneous space \cite{Wall}. Take $\cG_X = {\mathbb R}^2\ltimes {\bf SL}(2, {\mathbb R})$ (with natural action of ${\bf SL}(2, {\mathbb R})$ on ${\mathbb R}^2$) consist of the upper triangular matricies with positive diagonal entries.\,
$\gamma_G = 1$ 
because of the semidirect product of K-amenable groups 
(Proposision \ref{Pr11}).\,\,\,\,\,\,\,\,\,$\blacksquare$
\end{enumerate} 
The other five locally homogeneous spaces 
$
{\mathbb S}^4,\,\, {\mathbb S}^2\times{\mathbb R}^2,\,\,
{\mathbb S}^2\times{\mathbb S}^2,\,\,
{\mathbb S}^2\times{\mathbb H}^2,\,\,
{\mathbb S}^3\times{\mathbb R}
$
in Thurston's list are not locally symmetric spaces.
\begin{remark}
Are Corollaries \ref{Cor2} and \ref{Cor3} still valid without the locally symmetric assumption on $K(\Gamma, 1)$? 
The answer on this interesting question was found in 
{\rm \cite{Carey}} where the interesting conjecture has been formulated without the locally symmetric assumption on three-manifolds: 
Suppose $K(\G, 1)$ is a connected, compact, three-manifold which is an Eilenberg-MacLane space with fundamental group $\Gamma$. Then for any multiplier 
$\sigma\in H^2(\Gamma, U(1))$ on $\Gamma$,\, 
$
K_j(C_r^*(\Gamma,\sigma)) \cong K_j(C_r^*(\Gamma)) \cong K^{j+1}(K(\Gamma, 1)),\,j=0,1.
$
\end{remark}
We can generalize this statement for the case of four-manifolds in terms of conjecture: 
\begin{conjecture} 
Let $K(\G, 1)$ be a connected, compact, four-dimensional manifold which is an Eilenberg-MacLane space with fundamental group $\Gamma$. For any multiplier $\sigma\in H^2(\Gamma, U(1))$ on $\Gamma$ one has
\footnote{
Probably the Dixmier-Douady invariant $\delta(\sigma) = 0$  for all $\sigma \in H^2(\Gamma, U(1))$ for $\Gamma$ as in the conjecture.
}
\begin{equation}
K_j(C_r^*(\Gamma,\sigma)) \cong K_j(C_r^*(\Gamma)) \cong K^{j+1}(K(\Gamma, 1)), \quad j=0,1.
\end{equation}
\end{conjecture}

\section{T-duality and topology change from fluxes}

First we discuss T-duality in one direction only; then T-dualizing on the torus will be considered in the next sections. A more general case with T-dualizing on ${\mathbb T}^n,\, n>2$ can be obtained by appropriate succesive dualizations \cite{Bouwknegt1}. The application of T-duality is not restricted to product manifolds 
$M\times S^1$, but can also be applied locally in the case
of $S^1-$fibrations over $M$ \cite{Strominger}, and can be 
generalized to situations with nontrivial NS three-form 
flux $H$. 
A more general case where $X$ is an oriented $S^1-$bundle over the manifold $M$ characterizes by its first Chern class $c_1(X)\in H^2(M,\mathbb Z)$ in the presence of (possibly nontrivial) $H-$flux $\delta(B)\equiv [H]\in H^3(X,\mathbb Z)$. 
\footnote{Let us consider a bundle whose fiber $F$ is $(p-1)-$connected;
this means that for $k<p$ the $k-$th homotopy group 
$\pi_{k< p}(F)$ of $F$ vanishes. Let the $p$th homotopy group be nontrivial and is equal to the group $G$, 
$\pi_{k=p}(F) = G$. Then $F$ bundles can be characterized by a degree $(p+1)$ characteristic class in the cohomology with coefficients in $G$, $\omega_{p+1}\in H^{p+1}(X: G)$ and some characteristic classes of higher degree. Let all of the homotopy classes of $F$ of degree higher than $p$ vanish, then the bundle are characterized by $\omega_{p+1}$. In particular, for ${\mathbb Z}_2$ bundles $p=0$, $G= {\mathbb Z}_2$ and $\omega_1\in H^1(X; {\mathbb Z}_2)$ is the 
{\it spin} structure. Circle bundles, $p=1,\, G= {\mathbb Z}$,
characterized by a single characteristic class $c_1=\omega_2\in H^2(X; {\mathbb Z})$ which is called {\it the first Chern class}.
}
\begin{remark}
To simplify notations we will use the same notation for a cohomology class $[H]$, or for a representative $H$,
throughout the last part of this paper. For the reader it should be clear which is meant from the context.
\end{remark}
It has been argued \cite{Bouwknegt1} that
the T-dual of $X$, $\widehat{X}$, is again an oriented ${\widehat S}^1-$bundle over ${\widehat M}$
\begin{equation}
\begin{CD}
S^1 @>>> X \\
&& @V\pi VV \\
&& M \end{CD}
\,\,\,\,\,\,\, \stackrel{{\rm T-Duality}} {\longleftarrow\longrightarrow} \,\,\,\,\,\,\,
\begin{CD}
{\widehat S}^1 @>>> {\widehat X} \\
&& @V{\widehat \pi} VV \\
&& {\widehat M} \end{CD}
\end{equation}
supporting $H-$flux ${\widehat{H}}\in H^3({\widehat{X}}, 
{\mathbb Z})$, and $c_1(X) = \pi_* H$, 
$c_1({\widehat X}) = \widehat \pi_* \widehat H$.
Here $\pi_*  : H^k(X,{\mathbb Z})\to H^{k-1}(M,{\mathbb Z})$, and similarly $\widehat\pi_*$, denotes the pushforward maps. (As an example, at the level of the de Rham cohomology the pushforward maps $\pi_*$ and $\widehat \pi_*$ are simply the integrations along the $S^1-$fibers 
(${\widehat S}^1-$fibers ) of $X$ ($\widehat X$).)
\begin{proposition} (Gysin sequence)\label{PrB}
Let $\pi: X\rightarrow M$ be an oriented sphere bundle with fiber ${\mathbb S}^k$. Then there is a long exact sequence
(see for example {\rm \cite{Bott}})
\begin{equation}
\ldots \longrightarrow H^n(X)\stackrel{\pi_*}{\longrightarrow}
H^{n-k}(M)\stackrel{\wedge e}{\longrightarrow}
H^{n+1}(M)\stackrel{\pi^*}{\longrightarrow}
H^{n+1}(M)\longrightarrow \ldots
\end{equation}
where the maps $\pi_*, \wedge e$, and $\pi^*$ are an integration along the fiber, a multiplication by the Euler class, and the natural pullback, respectively.
\end{proposition}
In the case of an oriented $S^1$ bundle with first Chern class $c_1(X) = F\in H^2(M, {\mathbb Z})$ one gets
\begin{equation}
\ldots \longrightarrow H^\ell(M, {\mathbb Z}) \stackrel{\pi_*}{\longrightarrow}
H^{\ell}(X, {\mathbb Z})\stackrel{\pi_*}{\longrightarrow}
H^{\ell-1}(M, {\mathbb Z})\stackrel{F\cup}{\longrightarrow}
H^{\ell+1}(M, {\mathbb Z})\longrightarrow \ldots
\end{equation} 
In particular, for a two-dimensional base manifold $M$, the Gysin sequence gives an isomorphism between $H^3(X, {\mathbb Z})$ and $H^2(M,{\mathbb Z})$, i.e. between
Dixmier-Douady classes on $X$ and line bundles on $M$.  
For example ${\mathbb S}^3$ manifold can be considered as an $S^1-$bundle over ${\mathbb S}^2$ by means of the Hopf fibration. T-duality in the absence of $H-$flux leads to ${\mathbb S}^2\times S^1$ manifold supported
by one unit of $H-$flux \cite{Bouwknegt1}.

{\bf Circle bundles on Riemannian surfaces}.\,
Let us consider twisted K-groups of circle bundles over
two-manifolds and their T-duals. It can be shown that in this case $K^0$ of each space is related to $K^1$ of its dual. This class of manifolds includes:
\begin{itemize}
%\begin{enumerate}[${-}\,\,\, $]
\item{} The familiar examples of NS5-branes\,. 
\item{} Three-dimensional Lens spaces\,. 
\item{} ${\mathbb N}il-$manifolds\,. 
%\end{enumerate}
\end{itemize}
The K-groups determined by the Atiyah-Hirzebruch spectral sequence. It is convenient to consider the
first differential $d_3=Sq^3+H$ of the sequence only.
If $H^{\rm even}(X,{\mathbb Z})$ and 
$H^{\rm odd}(X,{\mathbb Z})$ are the even and odd
cohomology classes of the manifold $X$, then
the twisted K-groups are
\begin{equation}
K^0(X,H)=\frac{\textup{Ker}(H\cup:H^{\rm even}\longrightarrow H^{\rm odd})}
{H\cup H^{\rm odd}(X,{\mathbb Z})}\,,\quad
K^1(X,H)=\frac{\textup{Ker}(H\cup:H^{\rm odd}\longrightarrow H^{\rm even})}
{H\cup H^{\rm even}(X,{\mathbb Z})}\,.
\end{equation}

The case of noncommutative D2-branes has been consider in 
\cite{Carey}. These branes can wrap two-dimensional manifolds, which in the presence of a constant $B-$field are described by noncommutative Riemann surface.
Let $M = \Sigma_g \equiv {\mathbb H}^2/\Gamma_g$ be a Riemann surface of genus $g$. We can specialize to the case when 
$G= {\mathbb R}^2$,\, $K = \{e\}$ and $g=1$, with $\Gamma_1$
being ${\mathbb Z}^2$. 
Let $\sigma\in H^2(\Gamma_\sigma, {\bf U}(1))$ be any multiplier on $\Gamma_\sigma$. The graded groups are given by 
\begin{equation}
Gr(K^0(\Sigma_g)) = 
\oplus_j E^{2j}_{\infty}(\Sigma_g)\,,\,\,\,\,\,\,\,\,\,\,\,\,
Gr(K^1(\Sigma_g)) = 
\oplus_j E^{2j+1}_{\infty}(\Sigma_g)\,.
\end{equation}
In two dimensions the Chern character is an isomorphism over the integers and therefore we get
\begin{equation}
K^0(\Sigma_g)\cong H^0(\Sigma_g, {\mathbb Z})\oplus
H^2(\Sigma_g, {\mathbb Z})\cong {\mathbb Z}^2\,,
\,\,\,\,\,\,\,\,\,\,\,\,
K^1(\Sigma_g)\cong H^1(\Sigma_g, {\mathbb Z})\cong
{\mathbb Z}^{2g}\,.
\end{equation}
Using Theorem \ref{main} we have
\begin{equation}
K_\ell (C^*_r(\Gamma_g))\cong K^\ell (\Sigma_g)\,,
\,\,\,\,\,\,\,\,\,\,\,\,
K_\ell (C^*_r(\Gamma_g, \sigma))\cong 
K_\ell (\Sigma_g, \delta(B_\sigma))\,,\,\,\,\,\,\,\,\,\,
\ell =0, 1\,.
\end{equation}
Here $B_\sigma = C(\Sigma_g, E_\sigma)$. Note that 
$E_\sigma$ is a locally trivial flat bundle of
$C^*-$algebras over $\Sigma_g$,
with fibre $\mathcal{K}$ ($\mathcal K$ are compact operators),
it has a Dixmier-Douady invariant $\delta(B_\sigma)$ which can be viewed as the obstruction to $B_\sigma$ being Morita
equivalent to $C(\Sigma_g)$ \cite{Carey}. It is evident that
$
\delta(B_\sigma) = \delta(\sigma)\in 
H^3(\Sigma_g, \mathbb{Z}) = 0.
$
Thus $B_\sigma$ is Morita equivalent to $C(\Sigma_g)$
and finally
\begin{eqnarray}
&&
K_j(C^*_r(\Gamma_g, \sigma)) \cong 
K_j(C^*_r(\Gamma_g)) \cong  K^j(\Sigma_g)\,,
\nonumber \\
&&
K^0(\Sigma_g) \cong H^0(\Sigma_g, {\mathbb Z})
\oplus H^2(\Sigma_g, {\mathbb Z})  \cong  
{\mathbb Z}^2\,, 
\,\,\,\,\,\,\,\,\,\,\,
K^1(\Sigma_g) \cong  H^1(\Sigma_g, {\mathbb Z})
 \cong  {\mathbb Z}^{2g}\,.
\end{eqnarray}

{\bf Three-cycles}. For a Riemann surface of genus $g$,  $H^2({\Sigma}_g, {\mathbb Z})= {\mathbb Z}$ and topologically circle bundles are classified by an integer $j$. 
The cohomology of the total space $X^3$ are 
\footnote{
Example \cite{Carey}: In the flat case and for the Euclidean group 
$G = \mathbb R^{2n}\rtimes {\bf SO}(2n)$, $K= {\bf SO}(2n)$,
and $\Gamma \subset G$ is a Bieberbach group, i.e. 
$\Gamma$ is a uniform lattice in $G$. 
Also a generalization of {\it noncommutative flat manifolds}
can be defined by regarding
$C^*(\Gamma, \sigma)$ as such an object. In addition $\sigma$ is any group two-cocycle on $\Gamma$, due to the fact that
$
K_*(C^*(\Gamma,\sigma)) 
\cong K^{*}(\Gamma\backslash G/K).
$
}
\begin{enumerate}[${\bullet}\,\, $]
\item{}
$j=0$ (trivial line bundle):
$H^0(X^3,{\mathbb Z})= {\mathbb Z}$,\,\,
$H^1(X^3,{\mathbb Z})={\mathbb Z}^{2g+1}$,\,\,
$H^2(X^3,{\mathbb Z})={\mathbb Z}^{2g+1}$,\,\,
$H^3(X^3,{\mathbb Z})={\mathbb Z}$;
\item{}
$j\neq 0$ (the Chern class equal to $j$):
$H^0({X^3},{\mathbb Z})={\mathbb Z}$,\,\,
$H^1(X^3,{\mathbb Z})={\mathbb Z}^{2g}$,\,\,
$H^2(X^3,{\mathbb Z})={\mathbb Z}^{2g}\oplus 
{\mathbb Z}_j$,\,\,
$H^3(X^3,{\mathbb Z})={\mathbb Z}$\,.
\end{enumerate}
Then untwisted ($H=0$) and twisted ($H=k$) K-groups are given by \cite{Bouwknegt1}: 
\begin{eqnarray}
K^0(X^3,{H=0})= H^0(X^3,{\mathbb Z})\oplus H^2(X^3,{\mathbb Z}) & = & 
\, \left\{
\begin{matrix} {\mathbb Z}^{2g+2}\,\,\,\,\,&\textup{if $j=0$}\,,\cr 
{\mathbb Z}^{2g+1}\oplus{\mathbb Z}_j\,\,\,\,\, &\textup{if $j\neq 0$}\,,
\end{matrix}
\right.\nonumber\\
K^1(X^3,{H=0})= H^1(X^3,{\mathbb Z})\oplus H^3(X^3,{\mathbb Z}) & = & 
\, \left\{
\begin{matrix} {\mathbb Z}^{2g+2}\,\,\,\,\,\,\,\,\,\,\,\,\,\,\,\,\,\,\,
&\textup{if $j=0$}\,,\cr 
{\mathbb Z}^{2g+1}\,\,\,\,\,\,\,\,\,\,\,\,\,\,\,\,\,\,\,
&\textup{if $j\neq 0$}\,.
\end{matrix}
\right. \nonumber \\
K^0(X^3,{H=k}) = H^2(X^3,{\mathbb Z}) & = &
\left\{
\begin{matrix} {\mathbb Z}^{2g+1}\,\,\,\,\,\,\,\,\,\,\,\, &\textup{if $j=0$}\,,\cr 
{\mathbb Z}^{2g}\oplus{\mathbb Z}_j\,\,\,\,\,\,\,\,\,\,\,\,&\textup{if $j\neq 0$}\,,
\end{matrix}
\right.\nonumber \\
\!\!\!\!\!\!\!\!\!\!\!\!\!\!\!\!\!\!
K^1(X^3,{H=k})= H^1(X^3,{\mathbb Z})\oplus H^3(X^3,{\mathbb Z})/kH^3(X^3,{\mathbb Z}) & = & 
\left\{
\begin{matrix} {\mathbb Z}^{2g+1}\oplus{\mathbb Z}_k\,\,\,\,\, &\textup{if $j=0$}\,,\cr 
{\mathbb Z}^{2g}\oplus{\mathbb Z}_k\,\,\,\,\, &\textup{if $j\neq 0$}\,.
\end{matrix}
\right.
\label{K-gr}
\end{eqnarray}
In fact T-duality is the interchange of $j$ and $k$.  
Results in the twisted K-groups $K^0(X^3,H)$ and $K^1(X^3,H)$ being interchanged, which corresponds to the
fact that RR fieldstrengths are classified by $K^0(X^3,H)$ in type IIA string theory and by $K^1(X^3,H)$ in IIB.  
This means that applying the isomorphism between the two K-groups one can find the new RR fieldstrengths from the old ones. Indeed one simply interchanges the ${\mathbb Z}^{2g}$ between $H^1$ and $H^2$ and the rest of the cohomology groups are swapped $H^0\leftrightarrow H^1,\ H^2\leftrightarrow H^3$.

{\bf Lens spaces}. Let us consider the case of a linking two-sphere. It gives an isomorphism of the twisted K-theories of Lens spaces $L(1,p)=S^{3}/{\mathbb Z}_p$ (which is the
Eilenberg-MacLane space, see Remark \ref{R1})
\cite{Bouwknegt1}:
\begin{equation}
K^\ell(L(1,j),\, H=k)\cong K^{\ell+1}(L(1,k),\, H=j) \,.
\end{equation}
Recall that $L(1,p)={\mathbb S}^{3}/{\mathbb Z}_p$ is the total space of the circle bundle over the 
two-sphere with Chern class equal to $p$ times the generator 
of $H^2({\mathbb S}^2, {\mathbb Z})\cong {\mathbb Z}$. In particular, $L(1,1) = {\mathbb S}^3$ and 
$L(1,0) = {\mathbb S}^2 \times S^1$. 
The Lens space $L(2,j)$ can be considered as the nonsingular quotient $X={\mathbb S}^5/{\mathbb Z}_j$ when 
$j\neq 0$ and $X= {\mathbb C}P^2\times S^1$ when $j=0$. Integral cohomology groups of the oriented Lens space $L(n, q)$ are \cite{Bott}:
\begin{eqnarray}
&&
H^*(L(n, q), {\mathbb Z}) =
\left\{
\begin{matrix} \!\!\!{\mathbb Z}\,\,\,\,\,\,\,\, {\rm in}\,\,\, 
{\rm dimension}\,\,\,0\,, \cr 
\,\,\, {\mathbb Z}_q\,\,\,\,\,\,\, {\rm in}\,\,\,{\rm dimensions}\,\,\, 2n\,, \cr
\!\!\!\!\!\!\!\!\!\!
\!\!\!\!\!\!\!0\,\,\,\,\,\,\,\,\,\,{\rm otherwise}\,,
\end{matrix}
\right.
\,\,\,\,\,\,\,\,\,
H^{0\leq p\leq 5}({\mathbb C}P^2\times S^1)={\mathbb Z}\,,
\\
&&
H^0(L(2,j),{\mathbb Z})=H^5(L(2,j),{\mathbb Z})
= {\mathbb Z}\,,
\,\,\,\,\,\,\,\,\,\,\,\,\,\,\,\,\,\,
\,\,\,\,\,\,\,\,\,\,\,\,
H^2(L(2,j),{\mathbb Z})=H^4(L(2,j),{\mathbb Z})
= {\mathbb Z}_j \,.
\nonumber
\end{eqnarray}
A nontrivial quotient of string solution 
$AdS^5\times {\mathbb S}^5$  can be associated with circle bundles on ${\mathbb C}P^2$ 
(see for detail \cite{Bouwknegt1}).
It is easy to see that $H-$flux is only possible for the trivial bundle $j=0$, as the nontrivial bundles have trivial third cohomology. T-duality relates the trivial bundle with $H=j$ to the bundle with first Chern class $j$ and no flux
\cite{Bouwknegt1}.
\begin{remark}
The case $j=1$ of the T-duality mentioned above has been studied in {\rm \cite{Duff}}, and it was observed that the spacetime on the IIA side is not $Spin-$manifold, making the duality quite nontrivial. T-dualities considered in this section are interesting because IIB string theory on $AdS^5\times {\mathbb S}^5$ is well understood. Note that
the resulting RR fluxes are easily computed. Indeed, following the lines of {\rm \cite{Bouwknegt1}} we can start with $N$
units of $G_5-$flux supported on $L(2,j)$ in IIB. Then in IIA theory there will be $N$ units of $G_4-$flux supported on ${\mathbb C}P^2$ and $j$ units of
$H-$flux supported on $H^2({\mathbb C}P^2,{\mathbb Z})\otimes H^1(S^1,{\mathbb Z})$.
\end{remark}

\subsection{K-groups for branes and fluxes on 
${\mathbb R}P^7 \times {\mathbb T}^3$}
\begin{remark}
Note that factors ${\mathbb Z}^{2g}$ do not play important role in what follows. In fact, we can ignore them and consider the two-sphere $\Sigma_{g=0} = {\mathbb S}^2$ or two-torus $\Sigma_{g=1} = {\mathbb T}^2$ cases.
\end{remark}
\begin{proposition} \label{Pr1}
Let $M$ be a contractible. For freely acting of a group $G$ on $M$, $H_n(G, A) = H_n(X, A)$, and 
$H^n(G, A) = H^n(X, A)$, where $A$ is a trivial $G-$bundle
and $X\equiv G\backslash M$ is the orbit space $G$ in $M$, providing factor topology and canonic mapping $\pi$ from $M$
to $X$.
\end{proposition}
Let as before $G$ be a connected Lie group, $K$ be a maximal compact subgroup and $\Gamma$ be a closed discrete subgroup without torsion. (The $\Gamma-$acting on $G/K$ is defined by
$\gamma(gK)=(\gamma g)K$.) In accordance with Proposition \ref{Pr1} $H_n(\Gamma, A)= H_n(\Gamma\backslash G/K, A)$,\,
$H^n(\Gamma, A)= H^n(\Gamma\backslash G/K, A)$,\, where $A$ is a trivial $\Gamma-$module.
As an example, let $G= {\mathbb R}^n$,\, $\Gamma = {\mathbb Z}^n$, and
$K=\{0\}$. Then $X= {\mathbb T}^n$, and 
\begin{equation}
H^p({\mathbb T}^n, {\mathbb Z}) = 
H_p({\mathbb T}^n, {\mathbb Z})= {\mathbb Z}^
{\left(
\begin{matrix} n \cr 
p
\end{matrix}\right)}\,.
\end{equation}
The nontrivial classes for ${\mathbb T}^3$ are
\begin{eqnarray}
\begin{matrix} H^0({\mathbb T}^3, {\mathbb Z}) & = & 
H_0({\mathbb T}^3, {\mathbb Z}) = {\mathbb Z} \cr 
H^1({\mathbb T}^3, {\mathbb Z}) & = &
H_1({\mathbb T}^3, {\mathbb Z}) = {\mathbb Z}^3 \cr
H^2({\mathbb T}^3, {\mathbb Z}) & = &
H_2({\mathbb T}^3, {\mathbb Z}) = {\mathbb Z}^3 \cr
H^3({\mathbb T}^3, {\mathbb Z}) & = &
H_3({\mathbb T}^3, {\mathbb Z}) = {\mathbb Z}
\end{matrix}
\,\,\,\,\,\,\,\,\,\,\,\,\,\,\,\,\,
\begin{matrix}
K^0({\mathbb T}^3) & = & H^0({\mathbb T}^3, {\mathbb Z})\oplus
H^2({\mathbb T}^3, {\mathbb Z}) = {\mathbb Z}\oplus 
{\mathbb Z}^3 \cr
K^1({\mathbb T}^3) & = & H^1({\mathbb T}^3, {\mathbb Z})\oplus
H^3({\mathbb T}^3, {\mathbb Z}) = {\mathbb Z}\oplus
{\mathbb Z}^3 \cr
K_0({\mathbb T}^3) & = & H_0({\mathbb T}^3, {\mathbb Z})\oplus
H_2({\mathbb T}^3, {\mathbb Z}) = {\mathbb Z}\oplus
{\mathbb Z}^3 \cr
K_1({\mathbb T}^3) & = & H_1({\mathbb T}^3, {\mathbb Z})\oplus
H_3({\mathbb T}^3, {\mathbb Z}) = {\mathbb Z}\oplus
{\mathbb Z}^3
\end{matrix}
\end{eqnarray}
Torsion and (co)homology of ${\mathbb R}P^n$ are
\begin{eqnarray}
&&
\!\!\!\!\!\!\!\!\!\!\!\!\!\!
H^n({\mathbb R}P^n, {\mathbb Z}) =
\left\{
\begin{matrix} {\mathbb Z},\,\,\,\,\, {\rm if}\,\,\, n
\,\,\,\,\, {\rm odd} \cr 
{\mathbb Z}_2,\,\,\,\,\, {\rm if}\,\,\, n
\,\,\,\,\, {\rm even} 
\end{matrix}
\right.\,\,\,\,\,\,\,\,\,\,
H_n({\mathbb R}P^n, {\mathbb Z}) =
\left\{
\begin{matrix} {\mathbb Z},\,\,\,\,\, {\rm if}\,\,\, n
\,\,\,\,\, {\rm odd} \cr 
\,\,\,0,\,\,\,\,\, {\rm if}\,\,\, n
\,\,\,\,\, {\rm even} 
\end{matrix}
\right.
\nonumber \\
&&
\!\!\!\!\!\!\!\!\!\!\!\!\!\!
H^k({\mathbb R}P^n, {\mathbb Z}) = {\mathbb Z}, 0, 
{\mathbb Z}_2, 0, {\mathbb Z}_2, \ldots \,\,\,\,\,\,\,\,\,\,\,\,\,
H_k({\mathbb R}P^n, {\mathbb Z}) = {\mathbb Z}, 
{\mathbb Z}_2, 0, {\mathbb Z}_2, 0, \ldots\,,\,\,
k = 0, \ldots, n.
\end{eqnarray}
Therefore torsion, (co)homology and lower K-groups of ${\mathbb R}P^7$ become
\begin{eqnarray}
\begin{matrix} 
H_0({\mathbb R}P^7, {\mathbb Z}) & = & {\mathbb Z} \cr
H_1({\mathbb R}P^7, {\mathbb Z}) & = &{\mathbb Z}_2 \cr
H_2({\mathbb R}P^7, {\mathbb Z}) &= & 0 \cr
H_3({\mathbb R}P^7, {\mathbb Z}) &= &{\mathbb Z}_2 \cr
H_4({\mathbb R}P^7, {\mathbb Z}) &= & 0 \cr
H_5({\mathbb R}P^7, {\mathbb Z}) &= &{\mathbb Z}_2 \cr
H_6({\mathbb R}P^7, {\mathbb Z}) &= & 0 \cr
H_7({\mathbb R}P^7, {\mathbb Z}) &= &{\mathbb Z}
\end{matrix}
\,\,\,\,\,\,\,\,\,\,\,\,\,\,\,\,\,\,\,\,
\begin{matrix}
H^0({\mathbb R}P^7, {\mathbb Z}) & = & {\mathbb Z} \cr
H^1({\mathbb R}P^7, {\mathbb Z})& = &0 \cr
H^2({\mathbb R}P^7, {\mathbb Z})& = &{\mathbb Z}_2 \cr
H^3({\mathbb R}P^7, {\mathbb Z})& = & 0 \cr 
H^4({\mathbb R}P^7, {\mathbb Z})& = &{\mathbb Z}_2 \cr
H^5({\mathbb R}P^7, {\mathbb Z})& = & 0 \cr
H^6({\mathbb R}P^7, {\mathbb Z})& = &{\mathbb Z}_2 \cr
H^7({\mathbb R}P^7, {\mathbb Z})& = &{\mathbb Z} \cr
\end{matrix}
\,\,\,\,\,\,\,\,\,\,\,\,\,\,\,\,\,\,\,\,
\begin{matrix}
K_0({\mathbb R}P^7) & = & {\mathbb Z} \cr
K_1({\mathbb R}P^7) & = & {\mathbb Z}\oplus {\mathbb Z}_8 \cr
K^0({\mathbb R}P^7) & = & {\mathbb Z}\oplus {\mathbb Z}_8 \cr
K^1({\mathbb R}P^7) & = & {\mathbb Z} \cr
\end{matrix}
\end{eqnarray}
The K-theories of ${\mathbb R}P^7\times {\mathbb T}^3$  
can be combined in order to find the K-theory of the product using the K\"unneth formula
\begin{equation}
H_n(M_1 \times M_2)=\bigoplus_j \left( H_j(M_1)
\otimes H_{n-j}(M_2) \right) \bigoplus 
\left( \bigoplus_j {\rm Tor}(H_j(M_1), H_{n-j-1}
(M_2)) \right) \,.
\end{equation}
\begin{remark}
Let $M$ be a topological space (or finite CW-complex). 
Because of the K\"unneth formula one has $K_{\ell}(\torus^n) \cong K_{\ell}(S^1)^{\oplus 2^{n-1}}\cong 
{\mathbb Z}^{\oplus 2^{n-1}}$ for $\ell =0,1$. Thus the following isomorphisms are valid:
\begin{eqnarray} 
&&
\!\!\!\!\!\!\!
K_0(M \times {\mathbb T}^n)  \cong  \left(
\widetilde{K}_0 (M) \oplus
K_{1}(M) \oplus {\mathbb Z}\right)^{\oplus 2^{n-1}}
\!\!\!\!\!\!\!\!\!\!\!\!, \,\,\,\,\,\,\,\,
K_1 (M \times {\mathbb T}^n) \cong 
\left(K_{1}(M) \oplus
\widetilde{K}_0 (M) \oplus {\mathbb Z}\right)^{\oplus 2^{n-1}} 
\!\!\!\!\! \Longrightarrow
\nonumber \\
&&
\!\!\!\!\!\!\!
K_0 (M \times {\mathbb T}^n)  \cong  K_1 (M \times
{\mathbb T}^n) \ .
\label{Isom}
\end{eqnarray}
\end{remark}
The isomorphism (\ref{Isom}) is in fact the T-duality and describes a relationship between Type IIB and Type IIA D-branes on the spacetime $M\times {\mathbb T}^n$. This isomorphism exchanges wrapped D-branes
with unwrapped D-branes. In addition the powers of
$2^{n-1}$ give the expected multiplicity of D$p-$brane charges arising from wrapping all higher stable D-branes on various cycles of the torus ${\mathbb T}^n$.
\begin{remark}
For the compactification of type II (or type I) on an
$n-$torus ${\mathbb T}^n$ one can get the following isomorphisms {\rm \cite{Olsen}}
\footnote{
We use notation ${KO}$ for group $K_{\mathbb R}$.}:
\begin{eqnarray}
[K(KO)](M\times{\mathbb T}^n,{\mathbb T}^n) & = &\bigoplus_{k=0}^n
[\widetilde{K}(\widetilde{KO})]^{-k}(M)^{\oplus{n\choose k}}
\nonumber \\
& = &
[\widetilde{K} (\widetilde{KO})](M)^{\oplus 2^{n-1}}
\oplus [K (KO)]^{-1}(M)^{\oplus 2^{n-1}}
\nonumber \\
& \cong & 
[K (KO)]^{-1}(M\times{\mathbb T}^n,{\mathbb T}^n)\,.
\label{KTisos}
\end{eqnarray}
If $n=1$ then under the isomorphism (\ref{KTisos}) of 
K-groups, $\widetilde K(M)\otimes_{\mathbb Z}K({S}^1)$
maps to $\widetilde{K}(M)\otimes_{\mathbb Z}K^{-1}({S}^1)$ with the summands $K({S}^1)$ and
$K^{-1}({S}^1)$ interchanged. Thus, T-duality exchanges wrapped and unwrapped D-brane configurations. 
If $n>1$, then (\ref{KTisos}) gives the anticipated degeneracies $2^{n-1}$ of brane charges arising from the
higher supersymmetric branes wrapped on various cycles of the torus ${\mathbb T}^n$. 
\end{remark}
The result for ${\mathbb R}P^7\times {\mathbb T}^3$ is just the tensor product of the original K-theories because the K-theory of the torus has no torsion and so the ${\rm Tor}$ term in the K\"unneth formula is trivial
\begin{eqnarray}
&&
\!\!\!\!\!\!\!\!\!\!\!\!\!\!\!\!\!\!\!\!
K_0({\mathbb R}P^7\times {\mathbb T}^3)
= ((K_0({\mathbb R}P^7)\otimes K_0({\mathbb T}^3)) \oplus 
(K_1({\mathbb R}P^7)\otimes K_1({\mathbb T}^3)) = 
{\mathbb Z}^2\oplus {\mathbb Z}^4 \oplus {\mathbb Z}^2_8 \,, \nonumber\\
&&
\!\!\!\!\!\!\!\!\!\!\!\!\!\!\!\!\!\!\!\!
K_1({\mathbb R}P^7\times {\mathbb T}^3)
= ((K_0({\mathbb R}P^7)\otimes K_1({\mathbb T}^3)) \oplus 
(K_1({\mathbb R}P^7)\otimes K_0({\mathbb T}^3)) = 
{\mathbb Z}^2\oplus {\mathbb Z}^4 \oplus {\mathbb Z}^2_8 \,,
\nonumber\\
&&
\!\!\!\!\!\!\!\!\!\!\!\!\!\!\!\!\!\!\!\!
K^0({\mathbb R}P^7\times {\mathbb T}^3)
= ((K^0({\mathbb R}P^7)\otimes K^0({\mathbb T}^3)) \oplus 
(K^1({\mathbb R}P^7)\otimes K^1({\mathbb T}^3)) = 
{\mathbb Z}^2\oplus {\mathbb Z}^4 \oplus {\mathbb Z}^2_8 \,,
\nonumber\\
&&
\!\!\!\!\!\!\!\!\!\!\!\!\!\!\!\!\!\!\!\!
K^1({\mathbb R}P^7\times {\mathbb T}^3)
= ((K^0({\mathbb R}P^7)\otimes K^1({\mathbb T}^3)) \oplus 
(K^1({\mathbb R}P^7)\otimes K^0({\mathbb T}^3)) = 
{\mathbb Z}^2\oplus {\mathbb Z}^4 \oplus {\mathbb Z}^2_8 \,.
\end{eqnarray}
The universal coefficient theorem yields to the K-cohomology groups
\begin{equation}
K^0({\mathbb R}P^7\times {\mathbb T}^3)
=
K^1({\mathbb R}P^7\times {\mathbb T}^3)
= {\mathbb Z}^2\oplus {\mathbb Z}^4 \oplus {\mathbb Z}^2_8\,.
\end{equation}
The untwisted K-theories associated with real projective spaces ${\mathbb R}P^{2k+1}$ can be obtained identically. It is known that in type IIA string theory D-branes are classified by a group $K_1$ while RR field strengths are classified by a group $K^0$. In case of 
${\mathbb R}P^7\times {\mathbb T}^3$ both groups are ${\mathbb Z}^2\oplus {\mathbb Z}^4\oplus {\mathbb Z}^2_8$.
Therefore D-branes are classified by 
$K_1({\mathbb R}P^7\times {\mathbb T}^3)$ and the physical interpretation of the generators and branes wraps the ${\mathbb T}^3$ and the ${\mathbb R}P^7$ are similar to the description given in \cite{Bouwknegt2}.

\subsection{T-dualizing on 
${\mathbb C}P^3\times \Sigma_g \times {\mathbb T}^2$}
{\bf{Untwisted K-theory for the case $M={\mathbb C}P^3\times \Sigma_g \times {\mathbb T}^2$}}.
Here we dualize the ${\mathbb R}P^7 \times X^3$ case with no $H-$flux.
Let the manifold $X^3$ admit free cycle actions. In particular, both manifolds $X^3$ and ${\mathbb R}P^7$ are circle bundles over space the Riemannian two-space $\Sigma_g$ and the complex projective spaces ${\mathbb C}P^3$ respectively
\begin{equation}
\begin{CD}
{S^1} @>>> X^3\\
&& @V {\pi} VV \\
&& \Sigma_g  \end{CD} 
\,\,\,\,\,\,\,\,\,\,\,\,\,\,\,\,\,\,\,\,\,\,\,\,\,\,\,
\,\,\,\,\,\,\,\,\,\,\,\,\,
\begin{CD}
{{\widehat S}^1} @>>> {\mathbb R}P^{7}\\
&& @V\widehat{\pi} VV \\
&& {\mathbb C}P^3  \label{rp7} \end{CD} 
\end{equation}
The nontrivial cohomology classes of the base spaces are
\begin{equation}
H^0({\mathbb C}P^3, {\mathbb Z})\, = \, 
H^2({\mathbb C}P^3, {\mathbb Z}) \, =  \,
H^4({\mathbb C}P^3, {\mathbb Z}) \, = \, 
H^6({\mathbb C}P^3, {\mathbb Z}) \, = \, {\mathbb Z}\,. 
\end{equation}
We can T-dualize both circle fibers following the lines of \cite{Bouwknegt1,Bouwknegt2}. Note that T-duality exchanges the integrals of the $H-$fluxes over the circle fibers with the Chern classes. Since the original $H-$flux vanishes the dual Chern classes also vanish. Therefore
the dual spacetime $\widehat{M}$ consists of the product of two trivial circle bundles over the original base
\begin{equation}
\widehat{M}= {\mathbb C}P^3\times \Sigma_g \times S^1_\alpha\times S^1_\beta\,,  
\end{equation}
where $\alpha$ and $\beta$ associated with the Chern classes
$c_1\in H^2(\Sigma_g, {\mathbb Z})$ and 
$c_1\in H^2({\mathbb C}P^3, {\mathbb Z})$ of the bundles over $\Sigma_g$ and ${\mathbb C}P^3$ respectively.
The dual $H-$flux, $\widehat H$, is the sum of two $H-$fluxes that integrate to the Chern classes.

{\bf{Twisted K-theory for the case ${\widehat M} =
{\mathbb C}P^3\times \Sigma_g \times S^1_\alpha
\times S^1_\beta $}}.
The twisted K-theory of $\widehat{M}$, as the theory which has been performed due to an even number of T-dualities, 
must agree with the untwisted K-theory of the space 
${\mathbb R}P^7\times X^3$.
The K\"unneth formula for K-homology is
\begin{equation}
0\longrightarrow K_*(M_1)\otimes K_*(M_2)\longrightarrow K_*(M_1\times M_2)\longrightarrow \textup{Tor}(K_*(M_1),K_*(M_2))\longrightarrow 0
\end{equation}
Homology of $X^3$ with coefficients in ${\mathbb Z}$ can be computed using the result Eq. (\ref{K-gr}) and the universal coefficient theorem. 
\begin{theorem} (Universal Coefficient Theorem) For any space $X$ and associated abelian group $G$ the following result holds:

$\cT 1.$\,\,\, The homology group of $X$ with coefficients in $G$ has a splitting: 
$$
\!\!\!\!\!\!\!\!\!\!\!\!\!\!\! H_p(X; G)\simeq H_p(X)\otimes G 
\oplus {\rm Tor}\, (H_{p-1}(X, G))\,.
$$

$\cT 2.$\,\,\,  The cohomology group of $X$ with coefficients in $G$ has a splitting: 
$$
H^p(X; G)\simeq {\rm Hom}\,(H_p(X), G)\oplus {\rm Ext}\,
(H_{p-1}(X), G)\,.
$$
\end{theorem}
\begin{remark}
The (splittings) isomorphisms given by the universal coefficient theorem are said to be unnatural isomorphisms. The following maps of exact sequences are natural:
\begin{equation*}
\begin{array}{ccccccccc}
0 & \longrightarrow & H_p(X)\otimes G  & \longrightarrow & H_p(X; G) & \longrightarrow & {\rm Tor}(H_{p-1}(X), G)  & \longrightarrow & 0\,
\\
0 & \longrightarrow & H^p(X, {\mathbb Z})\otimes G   & \longrightarrow & H^p(X; G)  & \longrightarrow & {\rm Tor}(H^{p+1}(X; {\mathbb Z}), G) & \longrightarrow & 0\,
\\
0 & \longleftarrow & {\rm Hom}(H_p(X), G) & \longleftarrow & H^p(X; G)
& \longleftarrow & {\rm Ext}(H_{p+1}(X), G) & \longleftarrow & 0\,
\end{array}
\end{equation*}
\end{remark}
For the integer cohomology ($G={\mathbb Z}$) we have the following result: 
\begin{corollary}
For any space $X$ for which $H_\ell(X)$ and
$H_{\ell-1}(X)$ are finite generated ${\mathbb Z}-$modules it follows 
\begin{equation}
H^\ell(X)\simeq {\cF}_\ell(X)\oplus {\rm Tor}(H_{\ell-1}(X))\,.
\end{equation} 
Here ${\cF}_\ell(X)$ and ${\rm Tor}(H_{\ell-1}(X))$ are the free and torsion parts of $H_\ell(X)$ and $H_{\ell-1}(X)$, respectively.
\end{corollary}
We can evaluate the homology of the product using the K\"unneth formula. It gives
\begin{eqnarray}
K_0({\mathbb R}P^7 \times X^3) & = &
\left(K_0({\mathbb R}P^7)\otimes K_0(X^3)\right)
\oplus \left( K_1({\mathbb R}P^7)\otimes K_1(X^3) \right)
\nonumber \\
& &\oplus \textup{Tor}\left(K_0({\mathbb R}P^7),K_1(X^3)\right)
\oplus\textup{Tor}\left(K_1({\mathbb R}P^7),K_0(X^3)\right)\,,
\nonumber\\
K_1({\mathbb R}P^7 \times X^3) & = &
\left(K_0({\mathbb R}P^7)\otimes K_1(X^3)\right)
\oplus \left( K_1({\mathbb R}P^7)\otimes K_0(X^3) \right)
\nonumber \\
& &\oplus \textup{Tor}\left(K_0({\mathbb R}P^7),K_0(X^3)\right)
\oplus\textup{Tor}\left(K_1({\mathbb R}P^7),K_1(X^3)\right)\,.
\end{eqnarray}
\begin{remark}
Note that 
${\rm Tor}(A,B)$ vanishes unless both $A$ and $B$ contain torsion components. If $m$ and $n$ are positive integers, then ${\rm Tor}({\mathbb Z}_m, {\mathbb Z}_n)= 
{\mathbb Z}_{(m, n)}$, where $(m, n)$ denotes the greatest common divisor of $m$ and $n$.
\end{remark}
Therefore we get the following result
\begin{eqnarray}
K_0({\mathbb R}P^7 \times X^3; H=0, j=0) & = &
\left({\mathbb Z}^{2g+2}\otimes {\mathbb Z}\right)
\oplus \left({\mathbb Z}^{2g+2}\otimes({\mathbb Z}\oplus {\mathbb Z}_8)\right) \oplus 0 \oplus 0 
\nonumber \\
& = & 
{\mathbb Z}^{2g+3}\oplus {\mathbb Z}_8\,,
\nonumber \\
K_0({\mathbb R}P^7\times X^3; H=0, j\neq 0) & = &
\left(({\mathbb Z}^{2g+1}\oplus {\mathbb Z}_j)\otimes {\mathbb Z}\right)\oplus \left({\mathbb Z}^{2g+1} \otimes({\mathbb Z}\oplus {\mathbb Z}_8)\right)
\oplus {\rm Tor}({\mathbb Z}_j, {\mathbb Z}_8)
\oplus 0 
\nonumber \\
& = &
{\mathbb Z}^{2g+2}\oplus {\mathbb Z}_{j}\oplus
{\mathbb Z}_{8}\oplus {\mathbb Z}_{(j, 8)}\,,
\nonumber \\
K_0({\mathbb R}P^7\times X^3; H=k, j=0) & = &
\left({\mathbb Z}^{2g+1}\otimes {\mathbb Z}\right)
\oplus \left(({\mathbb Z}^{2g+1}\oplus {\mathbb Z}_k)\otimes({\mathbb Z}\oplus {\mathbb Z}_8)\right) \oplus 0 \oplus 0 
\nonumber \\
& = &
{\mathbb Z}^{2g+2}\oplus {\mathbb Z}_{k}
\oplus {\mathbb Z}_{8}\oplus {\mathbb Z}_{(k, 8)}\,,
\nonumber \\
K_0({\mathbb R}P^7\times X^3; H=k, j\neq 0) & = &
\left(({\mathbb Z}^{2g}\oplus {\mathbb Z}_j)\otimes 
{\mathbb Z}\right)
\oplus \left(({\mathbb Z}^{2g}\oplus 
{\mathbb Z}_k)\otimes({\mathbb Z}\oplus {\mathbb Z}_8)\right) \oplus {\rm Tor} ({\mathbb Z}_{j}, {\mathbb Z}_{8}) \oplus 0 
\nonumber \\
& = & 
{\mathbb Z}^{2g+1}\oplus {\mathbb Z}_{j}
\oplus {\mathbb Z}_{k}\oplus {\mathbb Z}_{8}
\oplus {\mathbb Z}_{(j, 8)}
\oplus {\mathbb Z}_{(k, 8)}\,,
\nonumber \\
K_1({\mathbb R}P^7\times X^3; H=0, j=0) & = &
\left({\mathbb Z}^{2g+2}\otimes ({\mathbb Z}\oplus {\mathbb Z}_8)\right)
\oplus \left({\mathbb Z}^{2g+2}\otimes {\mathbb Z}
\right) \oplus 0 \oplus 0 
\nonumber \\
& = & 
{\mathbb Z}^{2g+3}\oplus {\mathbb Z}_8\,,
\nonumber \\
K_1({\mathbb R}P^7\times X^3; H=0, j\neq 0) & = &
\left(({\mathbb Z}^{2g+1}\oplus {\mathbb Z}_j)\otimes ({\mathbb Z}\oplus {\mathbb Z}_8)\right)
\oplus \left({\mathbb Z}^{2g+1}\otimes {\mathbb Z}\right)
\oplus \, 0 \,\oplus \,0 
\nonumber \\
& & = 
{\mathbb Z}^{2g+2}\oplus {\mathbb Z}_{j}
\oplus {\mathbb Z}_{8}\oplus {\mathbb Z}_{(j, 8)}\,,
\nonumber \\
K_1({\mathbb R}P^7\times X^3; H=k, j=0) & = &
\left({\mathbb Z}^{2g+1}\otimes ({\mathbb Z}\oplus {\mathbb Z}_8)\right)
\oplus \left(({\mathbb Z}^{2g+1}\oplus {\mathbb Z}_k)\otimes {\mathbb Z}\right) \oplus 0 \oplus {\rm Tor}({\mathbb Z}_k, {\mathbb Z}_8)
\nonumber \\
& & =  {\mathbb Z}^{2g+2}\oplus {\mathbb Z}_{k}
\oplus {\mathbb Z}_{8}\oplus {\mathbb Z}_{(k, 8)}\,,
\nonumber \\
K_1({\mathbb R}P^7\times X^3; H=k, j\neq 0) & = &
\left(({\mathbb Z}^{2g}\oplus {\mathbb Z}_j)\otimes ({\mathbb Z}\oplus {\mathbb Z}_8)\right)
\oplus \left(({\mathbb Z}^{2g}\oplus {\mathbb Z}_k)\otimes {\mathbb Z}\right) \oplus 0 \oplus {\rm Tor}({\mathbb Z}_k, {\mathbb Z}_8) 
\nonumber \\
& = & 
{\mathbb Z}^{2g+1}\oplus {\mathbb Z}_{j}
\oplus {\mathbb Z}_{k}\oplus {\mathbb Z}_{8}
\oplus {\mathbb Z}_{(j, 8)}
\oplus {\mathbb Z}_{(k, 8)}\,.
\end{eqnarray}
In the absence of Neveu-Schwarz (NS) flux, $K^1$ group
describes RR field strengths in IIB string theory, which we write locally as $G_p = dG_{p-1}$. It has been suggested that RR field strengths are classified by twisted K-theory. 
Field strengths satisfy 
$
d_3G_p = (Sq^3 + H\cup)G_p =0\,.
$
Here $Sq^3$ is the Steenrod squares that takes torsion $p-$classes to torsion $(p+3)-$classes. 
\footnote{
In the classical limit, type II supergravity, we can forget the flux quantization condition and look at real cohomology. In fact,
we can no longer see the $Sq^3$ term. This theory contains 
RR potentials $C_{p-1}$, a NS three-form $H$ and a  gauge-invariance 
$
C_{p-1}\rightarrow C_{p-1}+d\Lambda_{p-2}+H\wedge \Lambda_{p-4} \,, 
$
for any set of forms $\Lambda_k$. It follows that there are two natural field strengths 
$G_{p}=dC_{p-1}$ and
$
F_p=dC_{p-1}+H\wedge C_{p-3}\,.
$
In addition, $G_p$ is closed and $F_p$ is gauge-invariant.
} 
In the case of S-duality, a configuration in which the 
$G_3-$flux valued in 
${\mathbb Z}\subset H^3({\mathbb R}P^7\times X^3)$ is nonzero, we can find another configuration which corresponds to a class in the K-theory twisted by the original $G_3$.
   Let us consider dualizing of 
${\mathbb R}P^7\times {\mathbb T}^3$ with no $H-$flux, example from Section 5.1. In this case we are looking for a trivial $G-$bundle (see Proposition \ref{Pr1}), 
$\Sigma_{g=1}= {\mathbb T}^2$ and $X^3= {\mathbb T}^3$.
Then, T-dualizing the circles becomes
\begin{eqnarray}
&&
\!\!\!\!\!\!\!\!\!\!\!\!\!\!\!\!\!
\!\!\!\!\!\!\!\!\!\!\!\!\!\!\!\!
\begin{CD}
{\widehat S}^1 
%@>>> 
\longrightarrow {\mathbb R}P^7 \\
&& 
\!\!\!\!\!\!\!\!
@V
{\widehat \pi} VV \\
%&&  
\end{CD}
\,\,\,\,\,\,\,\,
\begin{CD}
{S}^1 \times
%@>>> 
{\mathbb T}^2 \\
&& 
\!\!\!\!\!\!\!\!\!\!\!\!\!
@V
{\pi} VV \\
%&& 
\end{CD}
\nonumber \\
&&
\!\!\!\!\!\!\!\!\!\!\!\!\!
{\mathbb C}P^3 \,\,\,
\times \,
{\mathbb T}^2 \times {\mathbb T}^2 
\,\,\stackrel{{\rm T-Duality}}{\longleftarrow\longrightarrow}\,\,
\underbrace{
{\mathbb C}P^3 \times \Sigma_g \times 
S^1_{\alpha} \times S^1_{\beta}
}_{H\neq 0}
\,\,
\stackrel{{\rm 2T-Dualities}}{\longleftarrow\longrightarrow}
\,\,
\underbrace{
{\mathbb R}P^7 \times {\mathbb T}^3}_{H = 0}
\end{eqnarray}
The twisted K-theory associated with space 
${\widehat M} = {\mathbb C}P^3\times \Sigma_g 
\times S^1_\alpha\times S^1_\beta $,
as we have performed an even number of T-dualities,
must agree with the untwisted K-theory of the original space ${\mathbb R}P^7\times X^3$:
\begin{eqnarray}
&&
\!\!\!\!\!\!\!\!\!\!\!\!\!\!\!\!\!
\!\!\!\!\!\!\!\!\!\!\!\!\!\!\!\!
\begin{CD}
{\widehat S}^1 
%@>>> 
\longrightarrow {\mathbb R}P^7 \\
&& 
\!\!\!\!\!\!\!\!
@V
{\widehat \pi} VV \\
%&&  
\end{CD}
\,\,\,\,\,\,\,\,
\begin{CD}
{S}^1 
%@>>> 
\rightarrow {X^3} \\
&& 
\!\!\!\!\!\!\!
@V
{\pi} VV \\
%&& 
\end{CD}
\nonumber \\
&&
\!\!\!\!\!\!\!\!\!\!\!\!\!
{\mathbb C}P^3 \,\,\,\,\,\,
\times \,\,\,\,\,\,
\Sigma_g \times {\mathbb T}^2 \,\,\stackrel{{\rm T-Duality}}
{\longleftarrow\longrightarrow}\,\,
\underbrace{
{\mathbb C}P^3 \times \Sigma_g \times 
S^1_{\alpha} \times S^1_{\beta}
}_{H\neq 0}\,\,
\stackrel{{\rm 2T-Dualities}}{\longleftarrow\longrightarrow}
\,\,
\underbrace{
{\mathbb R}P^7 \times {X^3}}_{H = 0}
\end{eqnarray}

\section{Appendix}

\subsection{Eilenberg-MacLane spaces}
\begin{definition}
Let $\Gamma$ be a discrete group. A based topological space
(or CW complex) $X$ is called an Eilenberg-MacLane space of type $K(\Gamma, n)$, where $n\geq 1$, if
$\Gamma_k(X)= \Gamma$ for $k=n$ and $\Gamma_k(X)= 0$ for 
$k\neq n$. It means that all the homotopy groups $\Gamma_k(X)$ are trivial except for $\Gamma_n(X)$, which is isomorphic to $\Gamma$.
\end{definition}
Note that $K(\Gamma, 0)$ is a CW complex with $\Gamma_0=\Gamma$ having contractible components. 
For $n=1$ and finitely generated group $\Gamma$ the spaces $K(\Gamma, 1)$ are well known:
$K({\mathbb Z}, 1) = S^1$, $K({\mathbb Z}_2, 1) = 
{\mathbb R}P^{\infty}$, and $K({\mathbb Z}_m, 1) = L_m^{\infty}$ for $m> 2$. 
For $n\geq 2$\, group $\Gamma$ must be abelian.
For any abelian group $\Gamma$ with $n\geq 2$ there exists an Eilenberg-MacLane space of type $K(\Gamma, n)$ which can be constructed as a CW complex.
\footnote{
The Whitehead theorem implies that there is a unique $K(\Gamma, n)$ space up to homotopy equivalence in the category of topological spaces of the homotopy type of a CW complex.} 
In general they are more complicated objects which play a fundamental role in the connection between homotopy and (co)homology.
\begin{remark} \label{R1}
Recall that the simplest examples of Eilenberg-MacLane spaces are:
\begin{enumerate}[${\cR} 1.$]
\item{} $K({\mathbb Z}, 2)\thicksim {\mathbb C}P^{\infty} = S^{\infty}/S^1$;\, $H^*({\mathbb C}P^{\infty}; {\mathbb Z})\simeq 
{\mathbb Z}_p[t]$,\,\,\, $t\in H^2({\mathbb C}P^{\infty}; {\mathbb Z})$\, for all primes $p \geq 2$.
\item{} $K({\mathbb Z}_{p^n}, 1)\thicksim S^{\infty}/{\mathbb Z}_{p^n} = {\rm lim}_{N\rightarrow \infty}
L_{p^n}^{2N+1}(1, 1, ..., 1)$. 
The general lens space $L_m^{2n-1}(q_1, ..., q_{n-1})$ of dimension $2n-1$, is defined as the orbit space of the sphere ${\mathbb S}^{2n-1}\subset {\mathbb C}^n$ under the action of the group ${\mathbb Z}_m$ given by 
$
(z^1, ..., z^n)\mapsto 
(e^{2\pi\sqrt{-1}/m}z^1, e^{2\pi\sqrt{-1}q_1/m}z^2, ...,
e^{2\pi\sqrt{-1}q_{n-1}/m}z^n )\,,
$ 
where each $q_\ell$ is relatively prime to $m$. With this action of ${\mathbb Z}_m$
we set $L_m^{2n-1}(q_1, ..., q_{n-1}) = 
{\mathbb S}^{2n-1}/{\mathbb Z}_m$.
In particular, $K({\mathbb Z}_2, 1) = {\mathbb R}P^2 
= {\rm lim}_{N\rightarrow \infty} {\mathbb R}P^N$. 
\item{} $\Gamma = F$ (a free group),\, $K(F, 1)\thicksim S^1\vee\ldots\vee S^1$ 
(a bouquet of circles). 
\end{enumerate}
\end{remark}
Any two Eilenberg-MacLane spaces of $K(\Gamma, n)$ are weakly homotopy equivalent. Also $K(\Gamma, n)$ is a homotopy commutative $H-$space. The following result holds:
\begin{equation}
K(\Gamma_1\times \Gamma_2, n) = K(\Gamma_1, n)\times K(\Gamma_2, n)\,,\,\,\,\,\,\,\,
\Omega(K(\Gamma, n)) = K(\Gamma, n-1)\,,
\end{equation}
where $\Omega(K(\Gamma, n))$ is the loop space relative to some base point. The second equation is essentially a consequence of the homotopy groups isomorphism 
$\pi_\ell(\Omega(X))\simeq \pi_{\ell+ 1}(X)$.
\begin{theorem} (Hurewicz Isomorphism Theorem) \label{hu}
For $n-$connected cell complex $C$ the groups $\pi_{n+1}(C)$
and $H_{n+1}(C; {\mathbb Z})$\, ($n>0$) are isomorphic. 
\end{theorem}
By Hurwitz's Theorem \ref{hu}, $H_n(K(\Gamma, n); 
{\mathbb Z})\simeq
\pi_n(K(\Gamma, n))\simeq \Gamma$. It follows that there is a canonical isomorphism
$
H^n(K(\Gamma, n); G)\simeq {\rm Hom}\,(\Gamma, G)\,,
$
where $G$ is any abelian group and ${\rm Hom}\,(\Gamma, G)$
denotes the additive group of homomorphisms from the abelian group $\Gamma$ to $G$.

{\bf Partial computing of 
$H^*(K(\Gamma, n), {\mathbb Q})$ ring}.
Let $\Gamma$ be a finitely generated abelian group and let
$
\Gamma = {\mathbb Z}\oplus\ldots
\oplus{\mathbb Z}\oplus G\,,
$
where $G$ is a finite group. Then, respectively, 
$
K(\Gamma, n)= K({\mathbb Z}, n)\times
\cdot\cdot\cdot \times K(\Gamma, n)\times K(G, n)\,.
$
Because of the K\"unneth formulae we get
\begin{equation}
H^*(K(\Gamma, n); {\mathbb Q}) = H^*(K(\mathbb Z, n); {\mathbb Q})\otimes \ldots \otimes H^*(K(\mathbb Z, n); {\mathbb Q})\otimes
H^*(K(G, n); {\mathbb Q})\,.
\label{K}
\end{equation}
Note that if integer cohomologies of a topological space are finite then its rational cohomologies are trivial; 
$H^*(K(G, n);{\mathbb Q})= H^*(pt; {\mathbb Q})$. Therefore we have to compute $H^*(K({\mathbb Z}, n); {\mathbb Q})$, neglecting the last term in Eq. (\ref{K}).

Let $\bf k$ be a field, $\Lambda_{\bf k}(x_1, ..., x_n)$ (${\rm dim}\, x = n$) external algebra generated by $x_1, ..., x_n$, i.e.
${\bf k}-$algebra with generators $x_\ell$ and relations $x_kx_\ell=-x_\ell x_k, \, x_\ell^2 = 0$. The dimension of this algebra is $2^m$ and its basis form monoms $x_{\ell_1}\ldots x_{\ell_s}$,\,
$1\leq \ell_1 < ... < \ell_s\leq m$. Finally, ${\bf k}[x_1, ..., x_m]$ denotes an algebra of polynomials with coefficients in ${\bf k}$.
\begin{theorem}
\begin{equation}
H^*(K({\mathbb Z}, n); {\mathbb Q})
=\left\{ \begin{array}{ll} 
\Lambda_{\mathbb Q}(x)\,\,\,\,\,\,\,\,\,\,\,
{\rm for}\,\,\,{\rm odd}\,\,\, n\,,
\\
{\mathbb Q}[x]\,\,\,\,\,\,\,\,\,\,\,\,\,\,\,\,
{\rm for}\,\,\,{\rm even}\,\,\, n\,.
\end{array} \right.
\end{equation}
The formula 
$H^*(K({\mathbb Z}, n); {\mathbb Q})
= \Lambda_{\mathbb Q}(x)\, ({\rm dim}\,x = n)$ has a simple meaning: $H^*(K({\mathbb Z}, n); {\mathbb Q}) =
H^*({\mathbb S}^n; {\mathbb Q})$. The formula
$H^*(K({\mathbb Z}, n); {\mathbb Q}) = {\mathbb Q}[x]
\, ({\rm dim}\,x = n)$ means that 
$H^\ell(K({\mathbb Z}, n); {\mathbb Q}) = {\mathbb Q}$
for $\ell = 0, n, 2n, ...,$\, $H^\ell(K({\mathbb Z}, n); {\mathbb Q}) = 0$ for other value of $\ell$ and
the element $x^q$, $0 \neq x\in H^n(K({\mathbb Z}, n); {\mathbb Q})$, generates $H^{q n}(K({\mathbb Z}, n); {\mathbb Q})$ over ${\bf k}$.
\end{theorem}
\begin{corollary}
Let ${\rm rank}\,\Gamma = r$, then
\begin{equation}
H^*(K(\Gamma, n); {\mathbb Q})
=\left\{ \begin{array}{ll} 
\Lambda_{\mathbb Q}(x_1, ... ,x_r)\,\,\,\,\,\,
{\rm for}\,\,\,{\rm odd}\,\,\, n\,,
\\
{\mathbb Q}[x_1, ..., x_r]\,\,\,\,\,\,\,\,\,\,\,
{\rm for}\,\,\,{\rm even}\,\,\, n\,.
\end{array} \right.
\end{equation}
\end{corollary}
Let us consider the Eilenberg-MacLane complexes $K(\Gamma, n)$
for $\Gamma = {\mathbb Z}_{p^n}$ and $p$ prime.
We are looking for all cohomological operations modulo an arbitrary prime $p$. Remind that cohomologies of the space
$K({\mathbb Z}_p, 1)= L_p^{\infty}$ are known:
$H^q(K({\mathbb Z}_p, 1); {\mathbb Z})= {\mathbb Z}_p$ for
even $q$, and $H^q(K({\mathbb Z}_p, 1); {\mathbb Z})= 0$ for odd $q$. For positive $q$ there is a multiplicative isomorphism:
\begin{equation}
H^*(K({\mathbb Z}_p, 1); {\mathbb Z})\cong {\mathbb Z}_p
[x]\,,\,\,\,\,\,\,\, x\in H^2(K({\mathbb Z}_p, 1); 
{\mathbb Z})\,.
\end{equation}
\begin{theorem} (A. T. Fomenko and D. B. Fuks {\rm \cite{Fomenko}})
For $0< q \leq n+4p-3$ the ring $H^ q(K({\mathbb Z}, n); {\mathbb Z}_p)$ isomorphic to 
$H^ q(K({\mathbb Z}_p, n-1); {\mathbb Z})$.
\end{theorem}
For further interesting examples of computing of the groups
$H^*(K(\Gamma, n), {\mathbb Q})$ we refer the reader to the book \cite{Fomenko}.

\subsection*{Acknowledgements}
The author would like to thank Prof. L. Bonora for useful discussions and also acknowledge the Conselho Nacional de Desenvolvimento Cient\'ifico e Tecnol\'ogico (CNPq) for a support. The author also would like to thank Prof. M. Sasaki for hospitality during the scientific program 'Gravity and Cosmology 2007' at the Yukawa Institute for Theoretical Physics, Kyoto University.

\end{document}